\documentclass[journal=jacsat,manuscript=article]{achemso}
\SectionNumbersOn
\setkeys{acs}{maxauthors=10, etalmode=truncate}

\usepackage{chemformula} 
\usepackage[T1]{fontenc} 

\newcommand{\be}{\begin{equation}}
\newcommand{\ee}{\end{equation}}
\newcommand{\bea}{\begin{eqnarray}}
\newcommand{\eea}{\end{eqnarray}}
\newcommand{\ei}{\end{itemize}}
\newcommand{\ben}{\begin{enumerate}}
\newcommand{\een}{\end{enumerate}}

\newcommand{\lc}{\left[}

\newcommand{\rc}{\right]}
\newcommand{\lp}{\left(}
\newcommand{\rp}{\right)}

\def\gsim{\mathrel{\rlap{\lower4pt\hbox{\hskip1pt$\sim$}}
    \raise1pt\hbox{$>$}}}       
\def\lsim{\mathrel{\rlap{\lower4pt\hbox{\hskip1pt$\sim$}}
    \raise1pt\hbox{$<$}}}

\newcommand{\rep}{\mathrm{rep}}

\newcommand{\draft}[1]{}

\def \c{{\bf c}} 
\def\lapprox{\lower .7ex\hbox{$\;\stackrel{\textstyle <}{\sim}\;$}}
\def\gapprox{\lower .7ex\hbox{$\;\stackrel{\textstyle >}{\sim}\;$}}

\usepackage{tabularx}
\newcolumntype{C}[1]{>{\centering\arraybackslash}p{#1}}

\newcounter{daggerfootnote}

\title[Spatially-Resolved Bandgap and   Dielectric Function in 2D Materials
  from EELS]{ Spatially-Resolved Bandgap and   Dielectric Function in 2D Materials
  from Electron Energy Loss Spectroscopy }

\author{Abel Brokkelkamp}
\affiliation[Kavli]{Kavli Institute of Nanoscience, Delft University of Technology, 2628CJ Delft, The
  Netherlands}
\altaffiliation{Equal contribution}
\author{Jaco ter Hoeve}
\affiliation[Nikhef]{Nikhef Theory Group, Science Park 105, 1098 XG Amsterdam, The
  Netherlands}
\alsoaffiliation[VU]{Physics and Astronomy, VU Amsterdam,
  1081 HV Amsterdam, The Netherlands}
\altaffiliation{Equal contribution}
\author{Isabel Postmes}
\affiliation[Kavli]{Kavli Institute of Nanoscience, Delft University of Technology, 2628CJ Delft, The
  Netherlands}
\altaffiliation{Equal contribution}
\author{Sabrya E. van Heijst}
\affiliation[Kavli]{Kavli Institute of Nanoscience, Delft University of Technology, 2628CJ Delft, The
  Netherlands}
\author{Louis Maduro}
\affiliation[Kavli]{Kavli Institute of Nanoscience, Delft University of Technology, 2628CJ Delft, The
  Netherlands}
\author{Albert V. Davydov}
\affiliation[NIST]{Materials Science and Engineering Division, National Institute of Standards and Technology, Gaithersburg,
  MD, 20899 USA}
\author{Sergiy Krylyuk}
\affiliation[NIST]{Materials Science and Engineering Division, National Institute of Standards and Technology, Gaithersburg,
MD, 20899 USA}
\author{Juan Rojo}
\affiliation[Nikhef]{Nikhef Theory Group, Science Park 105, 1098 XG Amsterdam, The
  Netherlands}
\alsoaffiliation[VU]{Physics and Astronomy, VU Amsterdam,
  1081 HV Amsterdam, The Netherlands}
\author{Sonia Conesa-Boj}
\affiliation[Kavli]{Kavli Institute of Nanoscience, Delft University of Technology, 2628CJ Delft, The
  Netherlands}
\email{s.conesaboj@tudelft.nl}

\keywords{Scanning Transmission Electron Microscopy,
  Electron Energy
  Loss Spectroscopy, Neural Networks, Machine
  Learning, van der Waals materials, Bandgap, Dielectric Function.}

\begin{document}

\begin{tocentry}

  \begin{center}
    \includegraphics[width=0.62\textwidth]{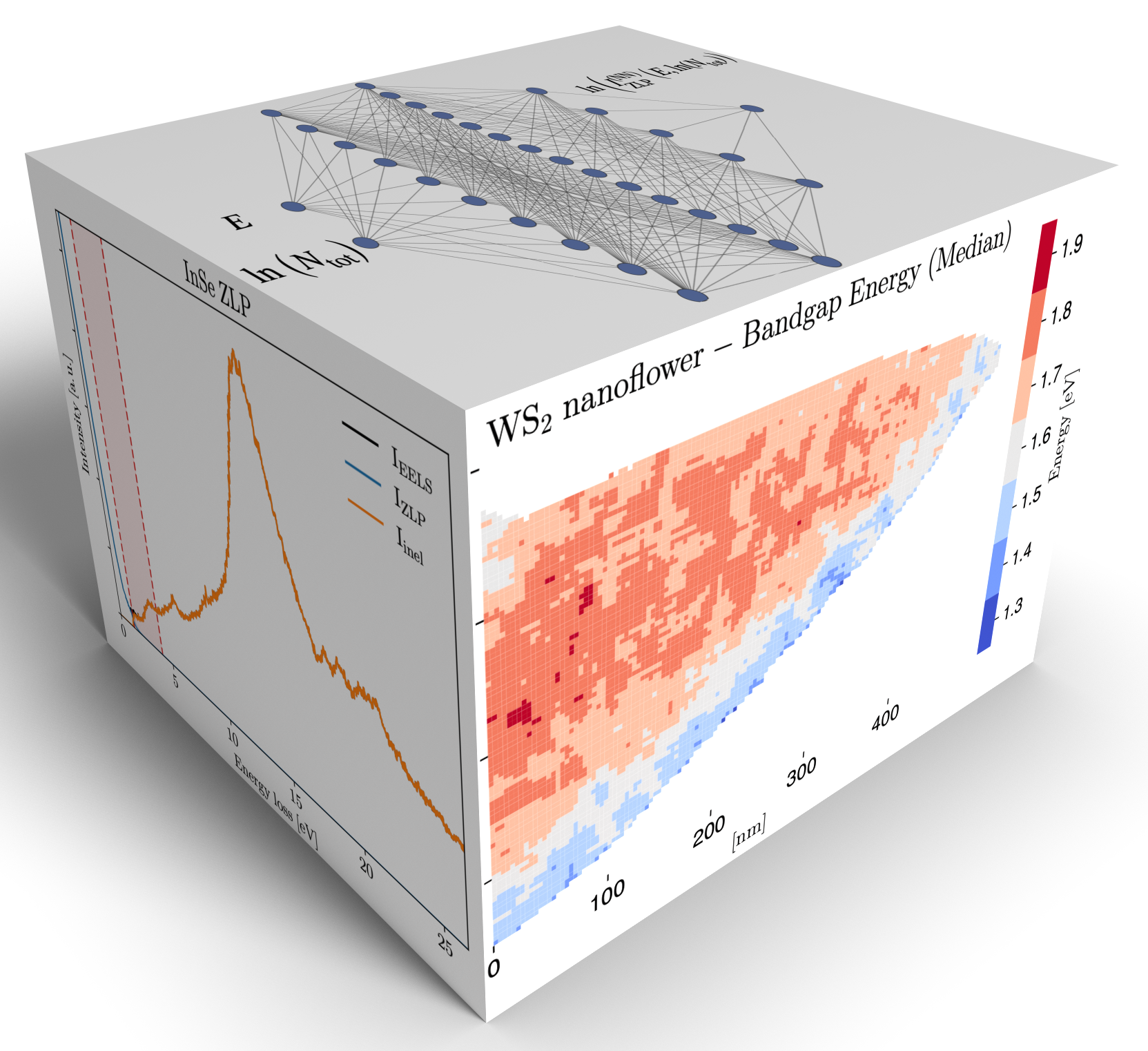}
  \end{center}

\end{tocentry}

\begin{abstract}
The electronic properties of two-dimensional (2D) materials depend sensitively on the underlying
atomic arrangement down to the monolayer level.
Here we present a novel strategy for
the  determination of the bandgap and complex
dielectric function in 2D materials
achieving a spatial resolution down to a few nanometers.
This approach is based on machine learning techniques developed
in particle physics and makes possible the automated processing
and interpretation
of spectral images from electron energy loss spectroscopy (EELS).
Individual spectra are classified as a function of the
thickness with $K$-means clustering and then used
to train a deep-learning model of the zero-loss peak background. 
As a proof-of-concept we assess the bandgap 
and  dielectric function
of InSe flakes and polytypic WS$_2$ nanoflowers,
and correlate
these electrical properties with the local thickness.
Our flexible approach is generalizable to other nanostructured materials
and to higher-dimensional spectroscopies,
and is made available as a new release of the open-source
{\sc\small EELSfitter} framework.
\end{abstract}

\section*{Introduction}
\label{sec:introduction}

Accelerating ongoing investigations of two-dimensional (2D) materials,
whose electronic properties depend on the underlying
atomic arrangement down to the single monolayer level, demands novel approaches able to map
this sensitive interplay with the highest possible resolution.
In this context, Electron Energy Loss Spectroscopy (EELS) analyses
in Scanning Transmission Electron Microscopy (STEM)
 provide access to a plethora of
structural, chemical, and local
electronic
information~\cite{Geiger:1967,Schaffer:2008,Erni:2005,Rafferty:1998,Stoger:2008},
from thickness and composition to the bandgap  and complex
dielectric function.
Crucially, EELS-STEM measurements can be
acquired as spectral images (SI), whereby each pixel
corresponds to a highly localised region of the specimen.
The combination of the excellent spatial and energy resolution
provided by state-of-the-art STEM-EELS
analyses~\cite{Terauchi:2005, Freitag:2005,Haider:1998} makes possible
deploying EELS-SI as a powerful and versatile tool to realise the
spatially-resolved simultaneous 
characterisation
of structural and electric properties in nanomaterials.
Such approach is complementary to related techniques such
as cathodoluminescence in STEM (STEM-CL), which however
is restricted to radiative processes while STEM-EELS probes both radiative
and non-radiative processes~\cite{Polman2019, doi:10.1021/acsphotonics.0c01950, Abajo:2010}. 

Fully exploiting this potential requires tackling two main challenges.
First, each SI is composed by up to tens of thousands of individual
spectra, which need to be jointly processed in a coherent manner.
Second, each  spectra is affected by a different Zero-Loss Peak (ZLP) background~\cite{Egerton:2009}, which depends
 in particular with the local thickness~\cite{Park:2008, Stoger:2008}.
A robust subtraction of this ZLP is  instrumental
to interpret the low-loss region, $E\lsim {\rm few~eV}$, in terms of
phenomena~\cite{Abajo:2010} such as 
phonons, excitons, intra- and inter-band transitions, and to determine
the local bandgap.
Furthermore, one should avoid the pitfalls of traditional
ZLP subtraction methods~\cite{Rafferty:2000, Egerton:1996,Dorneich:1998,Benthem:2001,Lazar:2003, EGERTON200247, Held:2020, Granerod:2018, Fung:2020} such as the need to specify an {\it ad hoc}
parametric functional dependence.

In this work we bypass these challenges
by presenting a novel strategy for
the  spatially-resolved determination of the bandgap and complex
dielectric function in nanostructured materials from
EELS-SI.
Our approach is based on machine learning (ML) techniques originally developed
in particle physics~\cite{Ball:2008by,Ball:2014uwa,Ball:2017nwa}
and achieves a spatial resolution down to a few nanometers.
Individual EEL spectra are first classified as a function of the
thickness with $K$-means clustering and subsequently used
to train a deep-learning model of the dominant ZLP background~\cite{Roest:2020kqy}.
The resultant ZLP-subtracted SI are amenable to theoretical
processing, in particular in terms of Fourier transform deconvolution
and Kramers-Kronig analyses, leading to a precise
determination of relevant structural and electronic properties at the nanoscale.

As a proof-of-concept we apply our strategy to the determination
of the bandgap 
and the complex dielectric function in two representative van der Waals materials,
InSe flakes
and polytypic WS$_2$ nanoflowers~\cite{WS2_nanoflowers}.
Both electronic properties are evaluated across the whole specimen and
can be correlated among them, e.g. to assess the interplay between bandgap
energy or the location of plasmonic resonances
with the local thickness.
Our approach is amenable to generalisation to other families
of nanostructured materials, is suitable for application
to higher-dimensional datasets such as momentum-resolved
EELS, and is made available as a new release of the 
{\sc\small EELSfitter} open-source framework~\cite{Roest:2020kqy}.

\section*{Computational Details}
\label{sec:methodology}

\begin{figure}[htbp]
    \centering
    \includegraphics[width=0.99\textwidth]{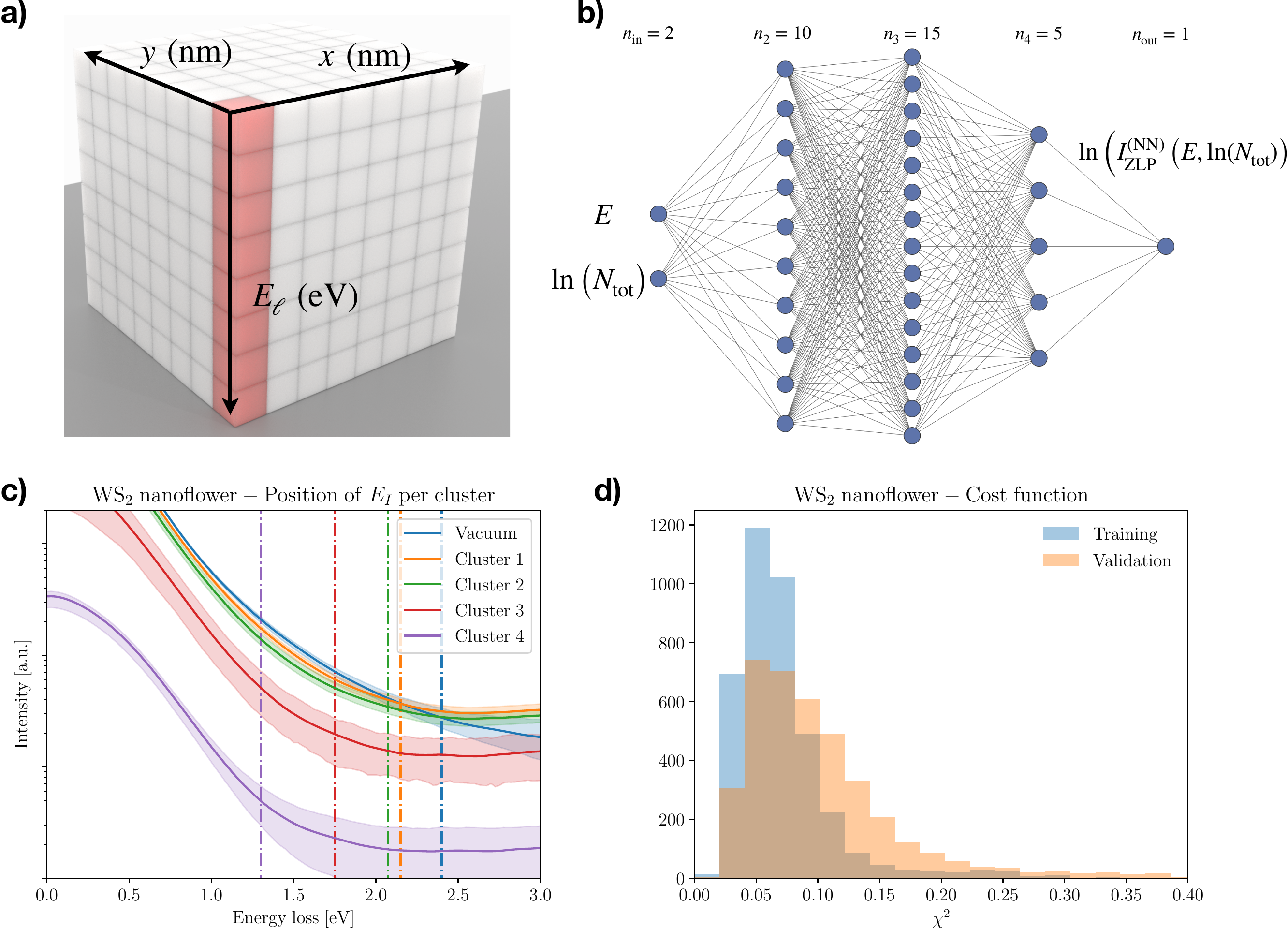}
    \caption{
(a) Schematic data-cube representing EELS-SI measurements, with two directions labeling
    the location across the specimen and the third one the energy loss,
    and whose entries are the total intensity $ I^{(i,j)}_{\rm EELS}(E_\ell)$ in
    Eq.~(\ref{eq:EELSmaster_image}).
      (b) The network architecture parametrising the ZLP.
      The input neurons are the energy loss $E$
      and the integrated intensity $N_{\rm tot}$,
      while the output neuron
      is the model prediction for the ZLP intensity.
      (c) The $E_{\rm I}$ hyperparameter defines the model training region,
      and is determined 
      from the first derivative $dI_{\rm EELS}/dE$ in each thickness cluster.
      (d) The training and validation cost function $C_{\rm ZLP}$,
      Eq.~(\ref{eq:costfunction_NNtraining}), evaluated
       over $5000$ models.
      Both (b) and (c) correspond to the WS$_2$ nanoflower specimen.
    }
    \label{fig:architecture}
\end{figure}

Spectral images in EELS-STEM are constituted by a large number, up to $\mathcal{O}(10^5)$,
of individual spectra
acquired across the analysed specimen.
They combine the excellent spatial resolution, $\mathcal{O}(40~{\rm pm})$,
achievable with STEM with the competitive
energy resolution, $\mathcal{O}(20~{\rm meV})$, offered by monochromated EELS.
From these EELS-SI it is possible to evaluate key
quantities such as the local thickness, the bandgap
energy and type, and the complex dielectric function, provided
one first  subtracts the  ZLP background which dominates the low-loss region of the EEL spectra.
The information provided by an EELS-SI can hence be represented by a three-dimensional data cube,
Fig.~\ref{fig:architecture}(a),
\begin{equation}
  \label{eq:EELSmaster_image}
  I^{(i,j)}_{\rm EELS}(E_\ell)= I^{(i,j)}_{\rm ZLP}(E_\ell) +
  I^{(i,j)}_{\rm inel}(E_\ell) \,, \quad i=1,\ldots, n_x\,, \quad j=1,\ldots, n_y\,,\quad  \ell=1,\ldots, n_E \, ,
\end{equation}
where $ I^{(i,j)}_{\rm EELS} $ indicates the total recorded  intensity
for an electron energy loss $E_\ell$ corresponding to the position $(i,j)$ in the specimen.
This intensity receives contributions from the inelastic scatterings
off the electrons in the specimen, $I_{\rm inel}$, and from the ZLP
 arising from elastic scatterings and instrumental broadening, $I_{\rm ZLP}$.
 In order to reduce statistical fluctuations, it is convenient
 to combine the information from neighbouring spectra using the
 pooling procedure described in the Supporting Information Sect.~S1.

Since the ZLP intensity depends strongly on the local  thickness of the specimen,
first of all we group individual spectra as a function of their thickness
by means of unsupervised machine learning, specifically by means of
the $K$-means clustering algorithm presented in Supporting Information Sect.~S1.
The cluster assignments are determined from the
minimisation of a cost function, $C_{\rm Kmeans}$, defined in thickness space,
\be
C_{\rm Kmeans}
= \sum_{r=1}^{n_x\times n_y}\sum_{k=1}^{K} d_{rk}\left|  \ln\lp \frac{\widetilde{N}^{(k)}}{N^{(r)}_{\rm tot}}\rp
\right|^p \, , \qquad r=i+(n_y-1)j \, ,
\ee
with $d_{rk}$ being a binary assignment variable, equal to 1 if $r$ belongs to cluster $k$
($d_{rk}=1$ for $r\in T_k$) and zero otherwise, and with the exponent satisfying $p> 0$.
Here $N^{(r)}_{\rm tot}$ represents the integral of $I^{(i,j)}_{\rm EELS}$ over the measured range
of energy losses, which provides a suitable proxy for the local thickness, and
$\widetilde{N}^{(k)}$ is the $k$-th cluster mean.
The number of clusters $K$ is a user-defined parameter.

Subsequently to this clustering, we train a deep-learning model
parametrising the specimen ZLP by extending the approach
that we developed in~\cite{Roest:2020kqy}.
The adopted neural network architecture is displayed in Fig.~\ref{fig:architecture}(b),
where the  inputs are the energy loss $E$ and the integrated intensity $ N_{\rm tot}$.
The model parameters ${\boldsymbol \theta}$
are determined from the minimisation of the cost function
\be
\label{eq:costfunction_NNtraining}
C_{\rm ZLP}({\boldsymbol \theta}) \propto \sum_{k=1}^K \sum_{\ell_k=1}^{n_E^{(k)}} \frac{\lc
  I^{(i_k,j_k)}(E_{\ell_k}) - I_{\rm ZLP}^{(\rm NN)} \lp
  E_{\ell_k},\ln \lp N_{\rm tot}^{(i_k,j_k)}\rp ;{\boldsymbol \theta}\rp \rc^2}{\sigma^2_k \lp E_{\ell_k}\rp } \, ,\qquad
E_{\ell_k} \le E_{{\rm I},k} \, ,
\ee
where within the $k$-th thickness cluster a representative spectrum $(i_k,j_k)$
is randomly selected, and with $\sigma_k \lp E_{\ell_k}\rp$
being the variance within this cluster.
The hyperparameters $E_{{\rm I},k}$ in Eq.~(\ref{eq:costfunction_NNtraining})
define the model training region for each cluster ($E_{\ell_k} \le E_{{\rm I},k}$)
where the ZLP dominates
the total recorded intensity.
They are automatically determined from the 
features of the first derivative $dI_{\rm EELS}/dE$,
e.g. by demanding that only $f\%$ of the replicas
have crossed $dI_{\rm EELS}/dE=0$, with $f\approx 10\%$.
Typical values of $E_{{\rm I},k}$ are displayed in Fig.~\ref{fig:architecture}(c),
where vacuum measurements are also included as reference.
To avoid overlearning, the input data is separated into disjoint training
and validation  subsets, with the latter used to determine
the optimal training length using look-back stopping~\cite{Ball:2014uwa}.
Fig.~\ref{fig:architecture}(d) displays
the distribution of the training and validation cost functions, Eq.~(\ref{eq:costfunction_NNtraining}), evaluated over $5000$ models.
Both Figs.~\ref{fig:architecture}(c) and~(d) correspond to
the WS$_2$ nanoflower specimen
first presented in~\cite{Roest:2020kqy} and revisited here.
Further details on the deep-learning model training are reported in Supporting Information Sect.~S2.

This procedure is repeated for a large number of models $N_{\rm rep}$,
each based on a different random selection
of cluster representatives, known
in this context as ``replicas''.
One ends up with a Monte Carlo representation
of the posterior probability density in the space of ZLP models, providing a faithful
estimate of the associated uncertainties,
\be
\label{eq:random_choice_spectra_v5}
{\boldsymbol I}_{\rm ZLP}^{(\rm NN)} \equiv \left\{ I_{\rm ZLP}^{({\rm NN})(n)} \lp
E,\ln \lp N_{\rm tot} \rp\rp \, , \quad  n=1,\ldots,N_{\rm rep}
\right\} \, ,
\ee
which makes possible a model-independent subtraction of the ZLP
and hence disentangling the contribution
from inelastic scatterings $I_{\rm inel}$.
Following a deconvolution procedure based in discrete Fourier transforms and reviewed in
Supporting Information Sect.~S3, these subtracted spectra allow us
to extract the single-scattering
distribution across the specimen and in turn the complex dielectric
function from a Kramers-Kronig analysis.
In contrast to existing methods, our approach provides an detailed estimate of the uncertainties
associated to the ZLP subtraction, and hence quantifies the statistical
significance of the determined properties by evaluating confidence level (CL) intervals
from the posterior distributions in the space of models.


\section*{Results and discussion}
\label{sec:results}

As a proof-of-concept we apply our strategy to two different 2D material specimens.
First, to horizontally-standing WS$_2$ flakes 
belonging to flower-like nanostructures (nanoflowers) characterised by a mixed 2H/3R polytypism.
This nanomaterial, member of the transition metal dichalcogenide (TMD)
family, was already considered in the original study~\cite{WS2_nanoflowers,Roest:2020kqy} and hence
provides a suitable benchmark to validate our new strategy.
One important property of WS$_2$ is that the indirect bandgap of its bulk form  switches
to direct  at  the monolayer level.
Second, to InSe nanosheets prepared by exfoliation of a Sn-doped InSe crystal
 and deposited onto a holey carbon TEM grid.
The electronic properties of InSe, such as the  band gap value and
type, are sensitive to both the layer stacking
($\beta$, $\gamma$, or $\varepsilon$-phase)
as well as to the magnitude and type of doping~\cite{GURBULAK2014106,JULIEN2003263,Rigoult:a18761,doi:10.1021/nn405036u}.
Supporting Information Sect.~S5 provides further details on the structural
characterisation of the InSe specimen.

\begin{figure}[t]
\begin{centering}
  \includegraphics[width=0.99\linewidth]{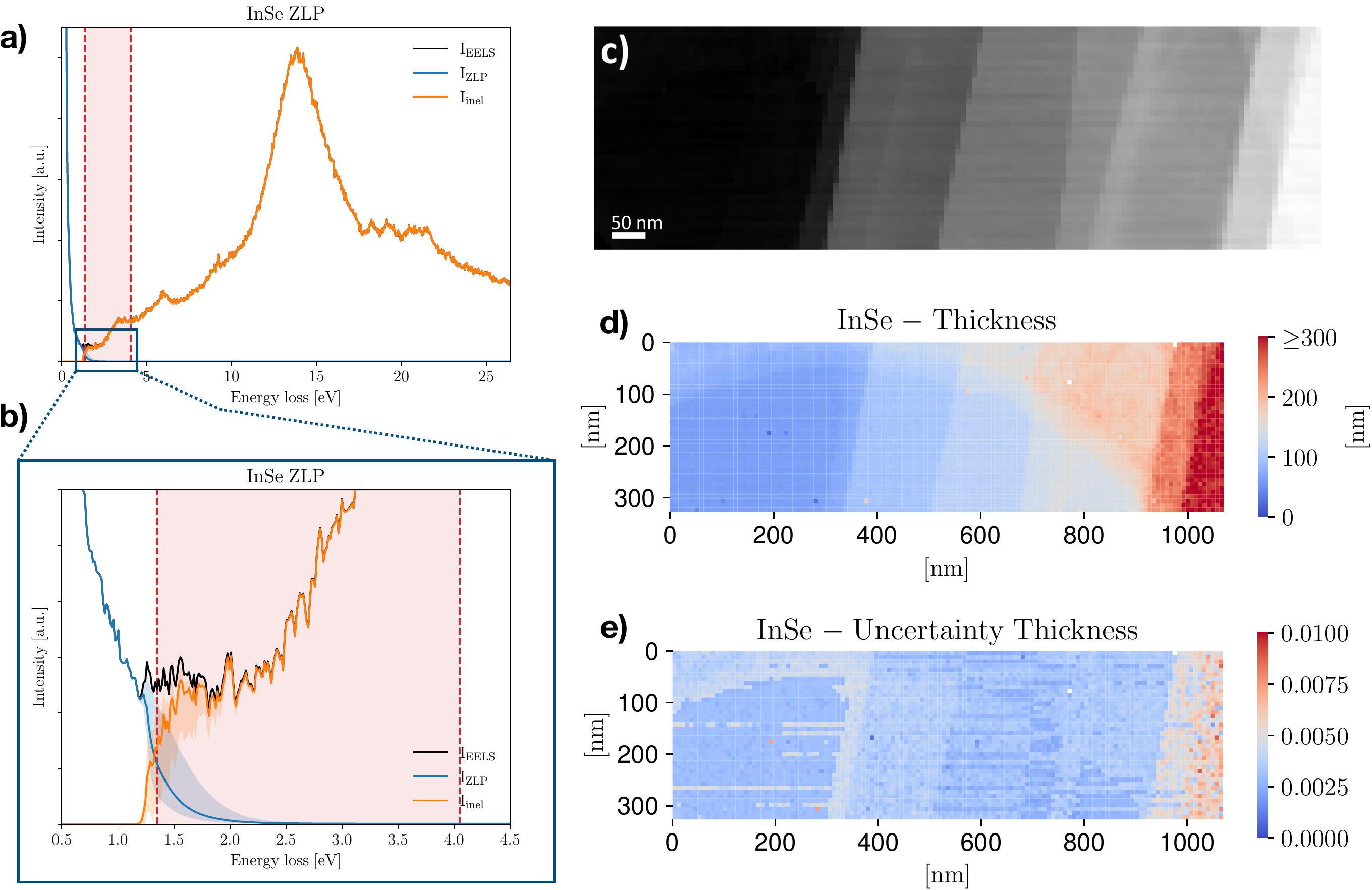}
  \caption{\small 
    (a) Representative EEL spectrum from the InSe specimen, where
    we display the data, the ZLP parametrization, and the subtracted
    inelastic spectrum.
    The red dashed region indicates the onset of inelastic scatterings where the bandgap is extracted.
    (b) Same spectrum, now zooming in the low-loss region marked with a blue square in (a).
    (c)  EELS-SI acquired on the InSe specimen displayed on Fig.~E.1(a,b) in the
    Supplementary Information, where each pixel
    corresponds to an individual spectrum.
    (d,e) The thickness map corresponding to the InSe SI of (c) and the associated relative uncertainties
    respectively.
  }
\label{fig:Fig2}
\end{centering}
\end{figure}

Fig.~\ref{fig:Fig2}(a) shows a representative EEL spectrum from
the InSe specimen, where the original data is compared
with the  deep-learning ZLP parametrisation and
the  subtracted inelastic contribution.
The red dashed region indicates the onset of inelastic scatterings, from which
the bandgap energy $E_{\rm bg}$ and type can be extracted from the procedure
described in Supporting Information Sect.~S4.
We zoom in Fig.~\ref{fig:Fig2}(b) in the low-loss region of the same
spectrum, where
the ZLP and inelastic components become of comparable size.
The error bands denote the 68\% CL intervals
evaluated over $N_{\rep}=5000$ Monte Carlo replicas.

By training the ZLP model on the whole InSe EELS-SI displayed
in Fig.~\ref{fig:Fig2}(c), see Fig.~E.1(a,b) in the
    Supplementary Information for
the corresponding STEM measurements,
we end up with a faithful
parameterisation of
$I_{\rm ZLP}^{(\rm NN)} \lp E, N_{\rm tot}\rp $ which can be used to
disentangle the inelastic contributions across the whole specimen
and carry out a spatially-resolved determination of relevant physical quantities.
To illustrate these capabilities, Fig.~\ref{fig:Fig2}(d,e)
displays the maps associated to the median thickness and its corresponding
uncertainties respectively
for the same InSe specimen, where a resolution of 8 nm is achieved.
One can distinguish the various terraces that compose the specimen, as well as the presence
of the hole in the carbon film substrate as a thinner semi-circular
region, see also the TEM analysis of Supporting Information Sect.~S5
The specimen thickness  is found to increase from around 20 nm to up to 300 nm as we move from left
to right of the map, while that of the carbon substrate is measured to be around 30 nm
consistent
with the manufacturer specifications.
Uncertainties on the thickness are below the 1\% level, as expected since its
calculation depends on the bulk (rather than the tails) of the ZLP.

\begin{figure}[t]
\begin{centering}
  \includegraphics[width=0.99\linewidth]{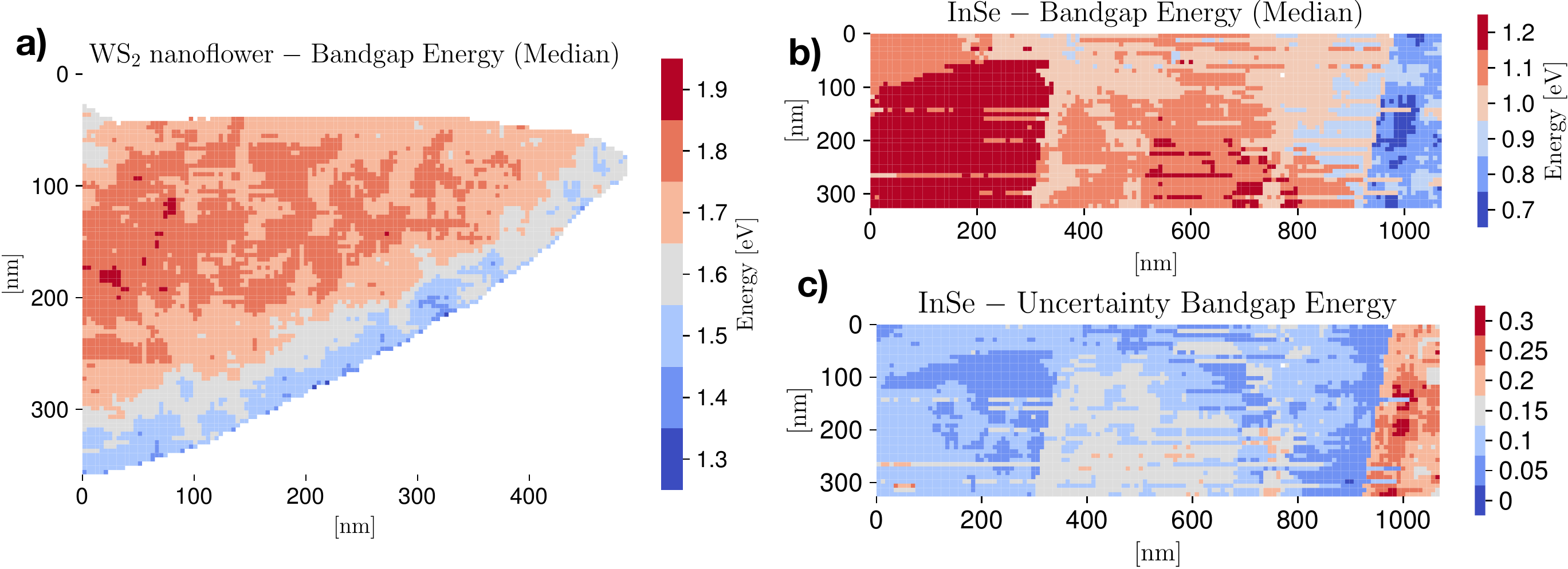}
  \caption{\small (a) Spatially-resolved map of the bandgap for the WS$_2$ nanoflower
    specimen, where a mask has been applied to remove the vacuum and pure substrate
    pixels.
    (b,c) The median value of the bandgap energy $E_{\rm bg}$ and its corresponding
    68\% CL relative uncertainties across the InSe specimen, respectively.
  }
\label{fig:Fig3}
\end{centering}
\end{figure}

In the same manner as for the thickness, the ZLP-subtracted SI contains
the required information
to carry out a specially-resolved determination of the bandgap.
For this, we adopt the approach of~\cite{Rafferty:1998}
where the behaviour of $I_{\rm inel}(E)$ in
the onset region is modeled as
\begin{equation}
    I_{\rm inel}(E)\simeq A(E-E_{\rm bg})^b,~~~~ E\gsim E_{\rm bg} \, ,
\end{equation}
where both the bandgap energy $E_{\rm bg}$ and the exponent $b$ are extracted from a fit to
the subtracted spectra.
The value of the exponent is expected to be around $b\approx 0.5~(\approx 1.5)$ for a semiconductor material
characterised by a direct (indirect) bandgap.
See Supporting Information Sect.~S4 for more details of this procedure.
Fig.~\ref{fig:Fig3}(a) displays the bandgap map for the
WS$_2$ nanoflower specimen, where a mask has been applied to remove the vacuum and pure-substrate
pixels.
A value $b=1.5$ for the onset exponent
is adopted, corresponding
to the reported indirect bandgap.
The uncertainties on $E_{\rm bg}$ are found to range between 15\% and 25\%.
The map of Fig.~\ref{fig:Fig3}(a)
is consistent with the findings of Ref.~\cite{Roest:2020kqy}~, which obtained
a value of the bandgap of 2H/3R polytypic WS$_2$ of $E_{\rm bg}=\lp 1.6\pm 0.3\rp $ eV
with a exponent of $b=1.3^{+0.3}_{-0.7}$  from a single  spectrum.
These results also
agree within uncertainties with first-principles calculations
based on Density Functional Theory for the band structure of 2H/3R polytypic
WS$_2$~\cite{doi:10.1021/acsphyschemau.1c00038}.
Furthermore, the correlation between the  thickness and bandgap maps 
points to a possible dependence of the value of $E_{\rm bg}$ on the specimen thickness,
though this trend is not statistically significant.
Further details about the bandgap analysis of the WS$_2$ nanoflowers
are provided in Supporting Information Sect.~S6.
    
Moving to the InSe specimen, Figs.~\ref{fig:Fig3}(b) and (c) display
the corresponding maps for the  median value of the bandgap energy and
for its uncertainties, respectively.
Photoluminescence (PL) measurements carried out on the same specimen,
{  and described in the Supporting Information Sect.~S5.},
indicate a direct bandgap with energy value around $E_{\rm bg}\approx 1.27$ eV,
hence we adopt $b=0.5$ for the onset exponent.
The median values of $E_{\rm bg}$ are found to lie in the range between 0.9 eV and 1.3 eV,
with uncertainties of 10\% to 20\% except for the thickest region where they
are as large as 30\%.
This spatially-resolved determination of the bandgap of InSe
is consistent with the spatially-averaged PL measurements as well
as with previous reports in the literature~\cite{henck2019evidence}.
Interestingly, there appears to be a dependence of $E_{\rm bg}$ with
the thickness, with thicker (thinner) regions in the right (left) parts
of the specimen favoring lower (higher) values.
This correlation, which remains robust once we account for the model
uncertainties, is
suggestive of the reported dependence of $E_{\rm bg}$  in InSe with the number of monolayers~\cite{Hamer}.

\begin{figure}[t]
\begin{centering}
  \includegraphics[width=0.99\linewidth]{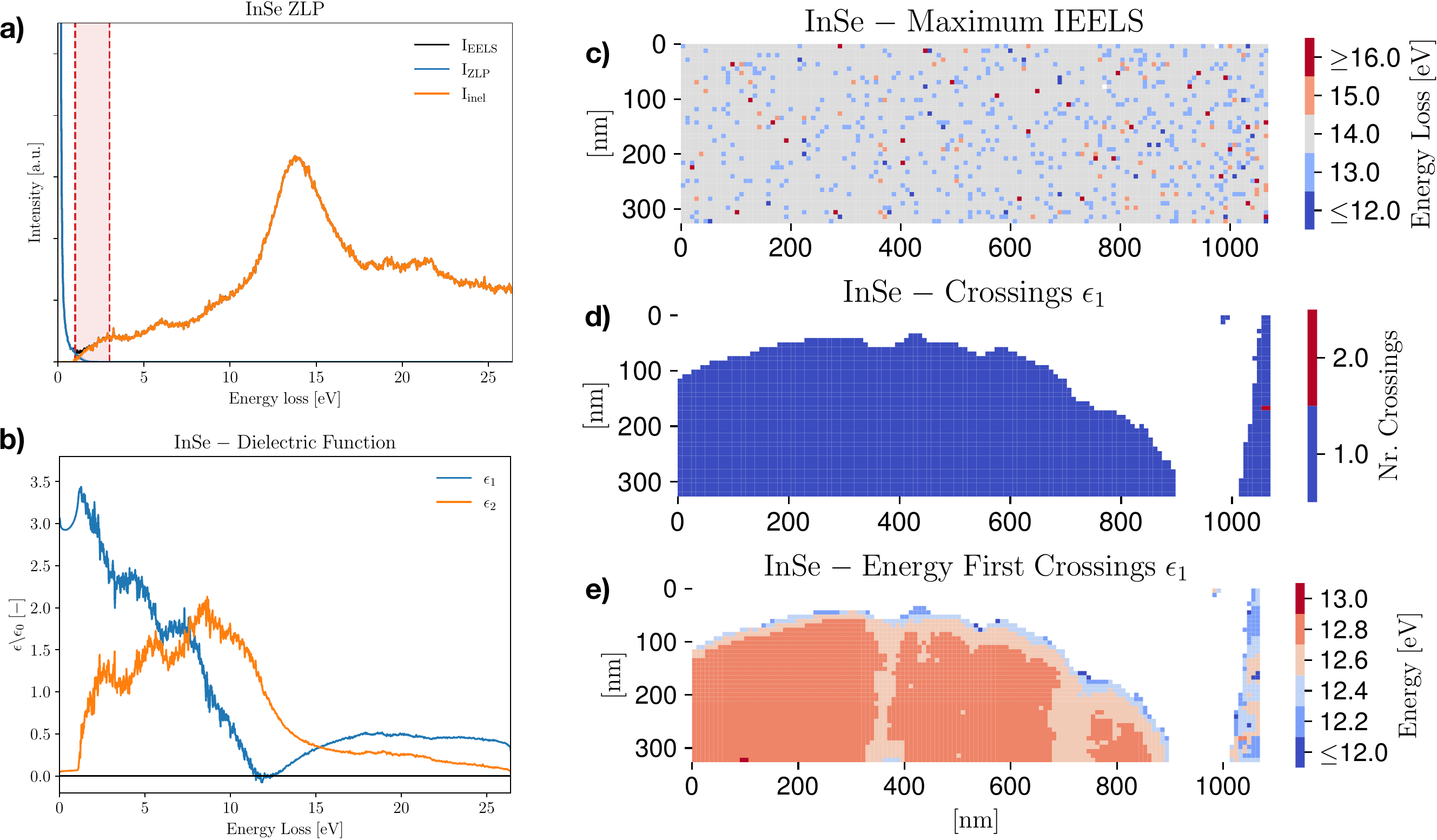}
  \caption{\small (a) A representative EEL spectrum from the InSe specimen.
    (b) The real, $\epsilon_1(E)$, and imaginary, $\epsilon_2(E)$,
    components of the complex dielectric function
    associated to the same location.
    (c) The energy value associated to the global maximum of the inelastic
    scattering intensity $I_{\rm inel}(E)$ across the InSe specimen.
    (d,e) The numbers of crossings of $\epsilon_1(E)$ and the associated
    value of the $E$ respectively across the same
    specimen, where the SI has been masked to
    remove pixels with carbon substrate underneath.
  }
\label{fig:Fig4}
\end{centering}
\end{figure}

{  Within our approach it is also possible
  to determine simultaneously the exponent $b$
  together with the bandgap energy $E_{\rm bg}$.
  As already observed in Ref.~~\cite{Roest:2020kqy}\,, this exponent
  is typically affected by large uncertainties.
  Nevertheless, it is found that in the case of the InSe specimen
  all pixels in the SI are consistent with $b=0.5$ and that the
  alternative scenario with $b=1.5$ is strongly disfavored.
  By retaining only those pixels where the determination
  of $b$ is achieved with a precision of better than 50\%, one finds
  an average value of
  $b = 0.50 \pm 0.26$,
  confirming that indeed this material is a direct semiconductor
  and in agreement with the spatially-integrated PL results.
  In addition, the extracted values of $E_{\rm bg}$ are found to be stable
  irrespectively of whether the exponent $b$ is kept fixed or instead  is
  also fitted.
  Supplementary Information Sect.~S8 provides more details on the
  joint $(E_{\rm bg},b)$ analysis. 
  }

We evaluate now the properties of the complex dielectric function $\epsilon(E)$
using the Kramers-Kronig analysis described in Supporting Information Sect.~S3.
In the following we focus on the InSe specimen, see Supporting Information Sect.~S7
for corresponding results for the WS$_2$ nanoflowers.
The local dielectric function provides key information on the  nature and location
of relevant electronic properties of the specimen.
To illustrate the adopted procedure,
Fig.~\ref{fig:Fig4}(a) displays another representative InSe spectrum
from the same EELS-SI of Fig.~\ref{fig:Fig3}(c).
Noticeable features include a marked peak at $E\approx 14$ eV, corresponding to the
bulk plasmon of InSe, as well as a series of smaller peaks in the low-loss region.
The real and imaginary parts of the complex dielectric function
associated to the same location in the InSe specimen are shown in Fig.~\ref{fig:Fig4}(b).
The values of the energy loss 
for which the real component exhibits a crossing, $\epsilon_1(E_c)=0$, with a positive slope
can be traced back to collective excitations such as a plasmonic resonances.
Indeed, one observes how the real component $\epsilon_1(E)$ exhibits a crossing
in the vicinity of $E\approx 13$ eV, consistent with the location of the bulk plasmon peak.

Furthermore, 
the local maxima of the imaginary component $\epsilon_2(E)$ can be associated to interband transitions.
From Fig.~\ref{fig:Fig4}(b), one finds that $\epsilon_2(E)$ exhibits local maxima in the low-loss
region, immediately after the onset of inelastic scatterings, at energy losses
around 3 eV, 6 eV, and
9 eV.
The location of these maxima do match with the observed peaks in the low-loss region of Fig.~\ref{fig:Fig4}(a),
strengthening their interpretation of interband transitions between
the valence and conduction bands, and consistent also
with previous reports in the literature~\cite{Politano}.
The dielectric function in Fig.~\ref{fig:Fig4}(b)
provides also access to $\epsilon_1(0)$,  the static dielectric constant
and hence  the refractive index $n$ of bulk InSe.
Our results are in agreement with  previous reports~\cite{https://doi.org/10.1002/pssb.2220960116}
once the thickness of our specimen is taken into account.

As for the thickness and the bandgap, one
can also map the variation of relevant
features in the dielectric function $\epsilon(E)$ across the specimen.
Extending the analysis of Figs.~\ref{fig:Fig4}(a,b),
Fig.~\ref{fig:Fig4}(c) shows the value of the energy loss associated to the
maximum of the inelastic
scattering intensity $I_{\rm inel}(E)$, while
Figs.~\ref{fig:Fig4}(d,e)  display the numbers of crossings of $\epsilon_1(E)$ and the
corresponding
value of the energy loss respectively.
In Figs.~\ref{fig:Fig4}(d,e), the SI has been masked to remove pixels with carbon substrate underneath, the reason being
that its contribution contaminates the recorded spectra and hence
prevents from robustly extracting $\epsilon(E)$ associated to InSe.
It is found that the specimen exhibits a single crossing whose energy $E_c$ ranges
between 12.5 eV and 13 eV, close to the  maximum of $I_{\rm inel}$ and hence
consistent with the location of the InSe bulk plasmonic resonance.
Uncertainties on $E_c$ are below the 1\% level,
since the calculation of  $\epsilon(E)$ depends mildly
on the onset region where model errors are the largest.
Dielectric function maps such as Fig.~\ref{fig:Fig4}(e) represent a sensitive method
to chart the local electronic properties of a nanostructured
material, complementing approaches such as
fitting multi-Gaussian models to EELS spectra to identify resonances and transitions.
In particular, maps for the local maxima of $\epsilon_1(E)$ and $\epsilon_2(E)$ could be also be constructed
to gauge their variation across the specimen.

Interestingly, as was also the case for the bandgap energy in Fig.~\ref{fig:Fig3}(c),
by comparing Fig.~\ref{fig:Fig4}(e) with Fig.~\ref{fig:Fig2}(d)
there appears to be a moderate correlation
between the crossing energy and the specimen thickness,
whereby $E_c$ decreases as the specimen becomes thicker.
While dedicated theoretical and modelling work would be
required to ascertain the origin of this sensitivity on the thickness,
our results  illustrate how our framework makes
possible a precise characterisation of the local electronic properties
of materials at the nanoscale and
their correlation with structural features.

\section*{Summary and outlook}
\label{sec:summary}

In this work we have presented a novel framework for the automated
processing and interpretation of spectral images in electron energy loss spectroscopy.
By deploying machine learning algorithms originally developed in 
particle physics, we achieve the robust subtraction of the ZLP background
and  hence a mapping of
the low-loss region in EEL spectra with precise spatial resolution.
In turn, this makes possible realising
a spatially-resolved ($\approx 10$ nm) determination of the  bandgap energy  and complex dielectric function in
layered materials, here represented by
2H/3R polytypic WS$_2$ nanoflowers and by InSe  flakes.
 We have also assessed how these electronic properties
 correlate with structural features, in particular with the local specimen thickness.
 Our results have been implemented in a new release of the
{\sc Python} open-source EELS analysis framework {\sc\small EELSfitter},
available from GitHub\footnote{ 
\url{https://github.com/LHCfitNikhef/EELSfitter}},
together with a detailed online documentation\footnote{
Available from \url{https://lhcfitnikhef.github.io/EELSfitter/index.html.}
}.

 While here we have focused on the interpretation of EELS-SI for layered
 materials, our approach is fully general
 and can be  extended both
to higher-dimensional datasets, such as  momentum-resolved EELS~\cite{MR-EELS} acquired in the
energy-filtered TEM  mode, as well as to different classes of nanostructured materials,
from topological insulators to complex oxides.
One could also foresee extending the method to the interpretation
of nanostructured materials stacked in heterostructures, and in particular to the removal of the substrate contributions, e.g.
for specimens fabricated on top of a solid substrate.
In addition, in this work we have restricted ourselves to a subset of the
important features contained in EEL spectra, while our approach could be extended
to the automated
identification and characterisation across
the entire specimen (e.g. in terms of peak position and width)
of the full range of plasmonic, excitonic, or intra-band transitions to streamline
their physical interpretation.
Finally, another exciting application of our approach would be to
assess the  capabilities of novel nanomaterials as prospective
light (e.g. sub-GeV) Dark Matter detectors~\cite{Knapen:2017xzo} by means
of their electron energy loss function~\cite{Knapen:2021run},
which could potentially extend the sensitivity of ongoing Dark Matter searches
by orders of magnitude.

\subsection*{Supporting Information}
\begin{itemize}
  \setlength\itemsep{-0.7em}
\item Technical details about the processing and
theoretical interpretation of EELS spectral images and
the ZLP subtraction.
\item Additional information about the determination of the bandgap
  energy and type as well as of the dielectric function.
  \item Details on the 
    structural characterisation of the InSe specimen, including
    PL measurements.
  \end{itemize}

\subsection*{Acknowledgments}

We are grateful to Irina Komen for carrying out the
photoluminiscence measurements.

\subsection*{Funding}

A~B. and S.~C.-B. acknowledge financial support
from the ERC through the Starting Grant ``TESLA”, grant agreement
no. 805021.
L.~M. acknowledges support from the
Netherlands Organizational for Scientific Research (NWO)
through the Nanofront program.
The work of J.~R. has been partially supported by NWO.
The work of J.~t.~H. is funded by NWO via a ENW-KLEIN-2 project.
S.~K. and A.~V.~D. acknowledge support through the
Materials Genome Initiative funding allocated to NIST.

Certain commercial equipment, instruments, or materials are identified in this paper
in order to specify the
experimental procedure adequately. Such identification is not intended to imply
recommendation or
endorsement by NIST, nor is it intended to imply that the materials or
equipment identified are necessarily the
best available for the purpose.

\subsection*{Declaration of competing interest}

The authors declare that they have no known competing financial interests or personal relationships that could have appeared to influence the work reported in this paper.

\subsection*{Methods}

\paragraph{STEM-EELS measurements.}
The STEM-EELS measurements corresponding to the WS$_2$ specimen 
were acquired with a JEOL 2100F microscope with a cold field-emission gun equipped with aberration corrector operated at 60 kV.
A Gatan GIF Quantum {
  ERS System (Model 966)} was used for the EELS analyses.
{
  The spectrometer camera was a Rio (CMOS) Camera.}
The convergence and collection semi-angles were 30.0 mrad and 66.7 mrad respectively.
{
  EEL spectra were acquired with an entrance aperture diameter of 5 mm, energy dispersion of   0.025 eV/ch, and exposure time of 0.001s.  For the STEM imaging and EELS analyses, a probe current of 18.1 pA and a camera length of 12 cm were used. EEL spectra size in pixels
  was a height  of 94 pixels and a  width of 128 pixels. }
The EELS data corresponding to the InSe specimen
were collected in a ARM200F Mono-JEOL microscope equipped with a GIF continuum spectrometer and operated at 200 kV.
{
  The spectrometer camera was a Rio Camera Model 1809 (9 megapixels).}
For these measurements, a slit in the monochromator of 1.3 $\mu$m was used.
A Gatan GIF Quantum {
  ERS System (Model 966)} was used for
the EELS analyses with
convergence and collection semi-angles of 23.0 mrad and 21.3 mrad respectively.
{
  EEL spectra were acquired with an entrance aperture diameter of 5 mm, energy dispersion of  0.015 eV/ch, and pixel time of  1.5 s. EEL spectra size in pixels
  was a height of 40 pixels and a width of 131 pixels.
%
}
For the STEM imaging and EELS analyses,
a probe current of {
  11.2 pA} and a camera length of 12 cm were used.

{ 
\paragraph{Photoluminiscence measurements.}
The optical spectra are acquired using a home-built spectroscopy set-up. The sample is illuminated through an 0.85 NA Zeiss 100x objective. The excitation source is a continuous wave laser with a wavelength of 595~nm and a power of 1.6 mW/mm$^2$ (Coherent OBIS LS 594-60). The excitation light is filtered out using colour filters (Semrock NF03-594E-25 and FF01-593/LP-25). The sample emission is collected in reflection through the same objective as in excitation, and projected onto a CCD camera (Princeton Instruments ProEM 1024BX3) and spectrometer (Princeton Instruments SP2358) via a 4f lens system. 
}

\clearpage

\appendix

\section{Pooling and clustering of EELS-SI}
\label{sec:processing-SI}

Let us consider a two-dimensional region
of the analysed specimen
with dimensions $L_x\times L_y$ where 
EEL spectra are recorded for $n_p=n_x \times n_y$ pixels.
Then the information contained within an EELS-SI may be expressed as
\begin{equation}
  \label{eq:EELSmaster_image_app}
  I^{(i,j)}_{\rm EELS}(E_\ell) \,, \quad i=1,\ldots, n_x\,, \quad j=1,\ldots, n_y\,,\quad  \ell=1,\ldots, n_E \, ,
\end{equation}
where $ I^{(i,j)}_{\rm EELS}(E_\ell) $ indicates the recorded total electron energy loss intensity
for an energy loss $E_\ell$ for a location in the specimen (pixel) labelled by $(i,j)$, and
$n_E$ is the number of bins that compose each spectrum.
The spatial resolution of the EELS-SI in the $x$ and $y$ directions is usually taken to be the same,
implying that
\be
\Delta x = \Delta y \approx \frac{L_x}{n_x} = \frac{L_y}{n_y} \, .
\ee
For the specimens analysed in this work we have $n_p=\mathcal{O}(10^4)$
spectra corresponding to a spatial resolution of $\Delta x \approx 10$ nm.
On the one hand, a higher spatial resolution is important to allow the identification and characterisation
of localised features within a nanomaterial,
such as structural defects, phase boundaries, surfaces or edges.
On the other hand, if the resolution
$\Delta x$ becomes too small the individual spectra become noisy due
to limited statistics.
Hence, the optimal spatial resolution can be determined from a compromise between these two considerations.

In general it is not known what the optimal spatial resolution should be prior to the STEM-EELS inspection
and analysis of a specimen.
Therefore, it is convenient to record the spectral image with a high spatial resolution and then, if required,
combine subsequently
the information on neighbouring pixels by means of a procedure known as pooling or
sliding-window averaging.
The idea underlying pooling is that one carries out the following replacement for the entries
of the EELS spectral image listed in Eq.~(\ref{eq:EELSmaster_image_app}):
\be
 I^{(i,j)}_{\rm EELS}(E_\ell) \quad \to  I^{(i,j)}_{\rm EELS}(E_\ell)\Big|_{\rm pooled} =\frac{1}{N^{(i,j)}_{\rm pool}}\sum_{|i'-i|\le d}
 \sum_{|j'-j|\le d}\lp \omega_{|i'-i|,|j'-j|} \times I^{(i',j')}_{\rm EELS}(E_\ell)\rp \, ,
\ee
where $d$ indicates the pooling range, $\omega_{|i'-i|,|j'-j|}$ is a weight factor, and
the pooling normalisation is determined by the sum of the relevant weights,
\be
N^{(i,j)}_{\rm pool} = \sum_{|i'-i|\le d} \sum_{|j'-j|\le d} \omega_{|i'-i|,|j'-j|} \, .
\ee
By increasing the pooling range $d$, one combines the local
information from a higher number of spectra and thus reduces statistical fluctuations,
at the price of some loss on the spatial resolution of the measurement.
For instance, $d=3/2$ averages the information contained on a $3\times 3$ square centered on the pixel
$(i,j)$.
Given that there is no unique choice for the pooling parameters, one has to verify that the interpretation
of the information contained on the spectral images does not depend sensitively on their value.
In this work, we consider uniform weights, $\omega_{|i'-i|,|j'-j|}=1$, but other options such as Gaussian weights
\be
\omega_{|i'-i|,|j'-j|} = \exp\lp - \frac{(i-i')^2}{2d^2} - \frac{(j-j')^2}{2d^2}  \rp \, ,
\ee
with $\sigma^2=d^2$ as variance are straightforward to implement in {\sc\small EELSfitter}.
The outcome of this procedure is  a  a modified spectral map 
with the same structure as Eq.~(\ref{eq:EELSmaster_image_app}) but now
with pooled entries.
In this work we typically use $d=3$ to tame statistical fluctuations
on the recorded spectra.

As indicated by Eq.~(\ref{eq:EELSmaster_image_app}), the total EELS intensity
recorded for each pixel of the SI receives contributions from both
inelastic scatterings and from the ZLP, where the latter must be subtracted before
one can carry out the theoretical interpretation of the low-loss region measurements.
Given that the ZLP arises from elastic scatterings with the atoms of the specimen,
and that the likelihood of these scatterings increases with the thickness, its contribution
will depend sensitively with the local thickness of the specimen.
Hence, before one trains the deep-learning model of the ZLP it is necessary to
first group individual spectra as a function of their thickness.
In this work this is achieved by means of unsupervised machine learning, specifically with
the $K$-means clustering algorithm.
Since the actual calculation of the thickness
has as prerequisite the ZLP determination, see Eq.~(\ref{eq:thickness_calculation}),
it is suitable to 
use instead the total integrated intensity as a proxy for the local thickness for the clustering procedure.
That is, we cluster spectra as a function of
\be
\label{eq:total_integrated_intensity}
N^{(i,j)}_{\rm tot} \equiv \int_{-\infty}^{\infty} dE\, I_{\rm EELS}^{(i,j)}(E) =
\int_{-\infty}^{\infty} dE\,\lp I^{(i,j)}_{\rm ZLP}(E) + I^{(i,j)}_{\rm inel}(E) \rp = N^{(i,j)}_0 + N^{(i,j)}_{\rm inel} \, ,
\ee
which coincides with the sum of the ZLP and inelastic scattering normalisation factors.
Eq.~(\ref{eq:total_integrated_intensity}) is inversely
proportional to the local thickness $t$ and therefore represents
a suitable replacement in the clustering algorithm.
In practice, the integration in Eq.~(\ref{eq:total_integrated_intensity}) is restricted to
the measured region in energy loss.

The starting point of  $K$-means clustering is a dataset composed by $n_p=n_x\times n_y$ points,
\be
\ln\lp N^{(r)}_{\rm tot}\rp \,,\quad r=1,\ldots , n_p\,, \qquad r=i+(n_y-1)j \, ,
\ee
which we want to group into $K$ separate clusters $T_k$, whose means are given by
\be
\ln \lp \widetilde{N}^{(k)}\rp \,,\quad k=1,\ldots, K\,.
\ee
The cluster means represent the main features of the $k$-th cluster
 to which the data points will be assigned in the  procedure.
 Clustering on the logarithm of $ N^{(r)}_{\rm tot}$ rather than on its absolute value
 is found to be more efficient, given that depending on the specimen location
 the integrated intensity will vary by orders of magnitude.

\begin{figure}[t]
\begin{centering}
  \includegraphics[width=0.95\linewidth]{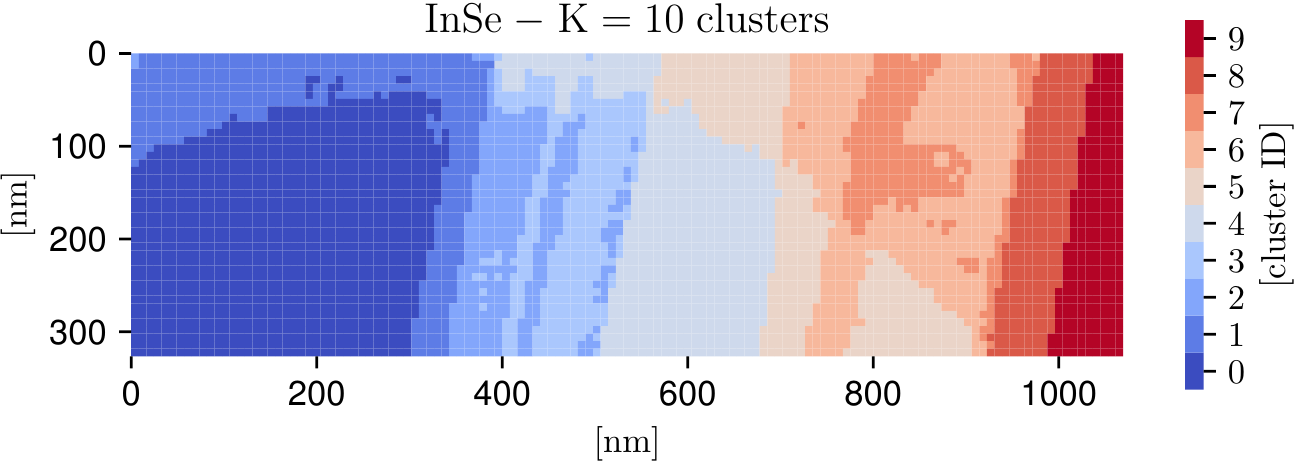}
  \caption{\small The outcome of the $K$-means clustering procedure
applied to the InSe specimen, where each color represents
one of the $K$=10 thickness clusters.
This clustering map can be compared with the
thickness map of Fig.~2(d), highlighting
how the total integrated intensity $N_{\rm tot}$ provides
a good proxy for the thickness.
  }
\label{fig:InSe_Clustered}
\end{centering}
\end{figure}
 
 In $K$-means clustering, the determination of the cluster means and data point
 assignments follows from the minimisation
of a cost function.
This is defined in terms of a distance in specimen thickness space, given by
\be
\label{eq:kmeans_clustering}
C_{\rm Kmeans}\lp {\boldsymbol N}_{\rm tot}, {\boldsymbol T}\rp
= \sum_{r=1}^{n_p}\sum_{k=1}^{K} d_{rk}\left|  \ln\lp \frac{\widetilde{N}^{(k)}}{N^{(r)}_{\rm tot}}\rp
\right|^p \, ,
\ee
with $d_{rk}$ being a binary assignment variable, equal to 1 if $r$ belongs to cluster $k$
($d_{rk}=1$ for $r\in T_k$) and zero otherwise, and with the exponent satisfying $p> 0$.
Here we adopt $p=1/2$, which reduces the weight of eventual outliers
in the calculation of the cluster means, and we verify that results
are stable if $p=1$ is used instead.
Furthermore, since clustering is exclusive, one needs to impose the following sum rule
\be
\sum_{k=1}^K d_{rk}=1 \, ,\quad \forall\,r \, .
\ee
The minimisation of Eq.~(\ref{eq:kmeans_clustering}) results in a cluster assignment such
that the internal variance is minimised and is carried out by means
of  a  semi-analytical  algorithm.
This algorithm is iterated until a convergence criterion is achieved, e.g. when the
change in the cost function between two iterations is below some threshold.
Note that,
as opposed to supervised learning, here is it not possible to overfit
and eventually one is guaranteed to find the solution that leads to the absolute minimum of
the cost function.
The end result of the clustering process is that now we can label the information contained
in the (pooled) spectral image (for $r=i+(n_y-1)j$) as follows
\be
\label{eq:cases_intensity}
  I^{(i,j)}_{{\rm EELS},k}(E_\ell) =
  \begin{cases}
                I^{(r)}_{\rm EELS}(E_\ell)\quad {\rm if} \quad r\in T_k  \\
              0 \quad {\rm otherwise}
            \end{cases} \,, \quad k=1,\dots, K\, .
  \ee
This cluster assignment makes
possible training  the ZLP deep-learning
model across the complete specimen recorded in the SI accounting
for the (potentially large) variations in the local thickness.

The number of clusters $K$ is a free parameter that needs to be fixed
taking into consideration how rapidly the local
thickness varies within a given specimen.
We note that $K$ cannot be too high, else it will  not  be possible
to sample a sufficiently large number of representative
spectra from each cluster to construct the prior probability
distributions, as required for the Monte Carlo
method used in this work.
We find that $K=10$ for the InSe and $K=5$ for the WS$_2$ specimens
are suitable choices.
Fig.~\ref{fig:InSe_Clustered} displays the
outcome of the $K$-means clustering procedure
applied to the InSe specimen, where each color represents
one of the $K$=10 thickness clusters.
It can be compared with the corresponding thickness map in
Fig.~2(d); the qualitative agreement further
confirms
that the total integrated intensity in each pixel $N^{(i,j)}_{\rm tot}$
represents a suitable proxy for the local specimen thickness.

\clearpage

\section{A deep-learning model for the zero-loss peak}
\label{app:nntraining}

Given that the zero-loss peak background cannot be evaluated
from first principles, in this work we deploy supervised machine learning
combined with Monte Carlo methods to construct a neural network
parameterisation of the ZLP.
Within this approach, one can faithfully model
 the ZLP dependence on both the
 electron energy loss and on the local specimen thickness.
Our approach, first presented in~\cite{Roest:2020kqy}, is here
extended to model the thickness dependence and to the simultaneous
interpretation of the $\mathcal{O}(10^4)$ spectra that constitute a typical EELS-SI.
One key advantage is the robust estimate of the uncertainties
associated to the ZLP modelling and subtraction procedures using
the Monte Carlo replica method~\cite{DelDebbio:2007ee}.

The neural network architecture adopted in this work is displayed
in Fig.~2(b).
It contains two input variables, namely
the energy loss $E$ and the logarithm of the integrated intensity $
\ln\lp N_{\rm tot}\rp$, the latter providing a  proxy for the thickness $t$.
Both $E$ and $\ln\lp N_{\rm tot}\rp$ are preprocessed and rescaled
to lie between 0.1 and 0.9 before given as input to the network.
Three hidden layers contain 10, 15, and 5 neurons respectively.
The activation state of the output neuron in the last layer,
$\xi^{(n_L)}_1$, is then related to the intensity of
the ZLP as
\be
I_{\rm ZLP}^{(\rm NN)}(E,\ln (N_{\rm tot}) ) = \exp\lp \xi^{(n_L)}_1( E, \ln ( N_{\rm tot})) \rp \, ,
\ee
where an exponential function is chosen to facilitate the learning, given that
the EELS intensities in the training dataset can vary by orders of magnitude.
Sigmoid activation functions are adopted for
all layers except for a ReLU in the final layer, to guarantee
a positive-definite output of the network and hence of
the predicted intensity.

The training of this neural network model for the ZLP is carried out
as follows.
Assume that the input SI  has been  classified into $K$ clusters
following the procedure of App.~\ref{sec:processing-SI}.
The members of each cluster exhibit a similar value of the local thickness.
Then one selects at random a representative spectrum from each cluster,
\be
\label{eq:random_choice_spectra_app}
\left\{ I_{\rm EELS}^{(i_1,j_1)}(E), I_{\rm EELS}^{(i_2,j_2)}(E),\ldots, I_{\rm EELS}^{(i_K,j_K)}(E) \right\} \, ,
\ee
each one characterised by a different total integrated intensity evaluated from
Eq.~(\ref{eq:total_integrated_intensity}),
\be
\left\{ N_{\rm tot}^{(i_1,j_1)}, N_{\rm tot}^{(i_2,j_2)},\ldots, N_{\rm tot}^{(i_K,j_K)} \right\} \, ,
\ee
such that $(i_k,j_k)$ belongs to the $k$-th cluster.
To ensure that the neural network model accounts only for 
the energy loss $E$ dependence in the region where the ZLP dominates the 
recorded spectra, we remove
from the training dataset those bins with $E \ge  E_{{\rm I},k}$
with $E_{{\rm I},k}$ being a model hyperparameter~\cite{Roest:2020kqy}
which varies in each thickness cluster.
The cost function $C_{\rm ZLP}$ used to train the NN model  is then
\be
\label{eq:costfunction_NNtraining_appendix}
C_{\rm ZLP} = \frac{1}{n_{E}}\sum_{k=1}^K \sum_{\ell_k=1}^{n_E^{(k)}} \frac{\lc
  I_{\rm EELS}^{(i_k,j_k)}(E_{\ell_k}) - I_{\rm ZLP}^{(\rm NN)} \lp
E_{\ell_k},\ln \lp N_{\rm tot}^{(i_k,j_k)}\rp \rp \rc^2}{\sigma^2_k \lp E_{\ell_k}\rp } \, ,\qquad
E_{\ell_k} \le E_{{\rm I},k} \, ,
\ee
where the total number of energy loss bins that enter the calculation
is the sum of bins in each individual spectrum,
$
n_{E} = \sum_{k=1}^K n_E^{(k)} \, .
$
The denominator of Eq.~(\ref{eq:costfunction_NNtraining_appendix}) is given by
$\sigma_k \lp E_{\ell_k}\rp$,
which represents the variance within the $k$-th cluster for a given value of
the energy loss $E_{\ell_k}$.
This variance is evaluated
as the size of the 68\% confidence level (CL) interval of the intensities
associated to the $k$-th cluster for a given value of $E_{\ell_k}$.

For such a random choice of representative cluster spectra, Eq.~(\ref{eq:random_choice_spectra_app}),
the parameters (weights and thresholds) of the neural network
model are obtained from the minimisation of Eq.~(\ref{eq:costfunction_NNtraining_appendix})
until a suitable convergence criterion is achieved.
Here this training is carried out using stochastic gradient descent (SGD)
as implemented in the {\tt PyTorch} library~\cite{NEURIPS2019_9015}, specifically
by means of the ADAM minimiser.
The optimal training length is determined by means of the look-back cross-validation
stopping.
In this method, the training data is  divided 80\%/20\%  into training
and validation subsets, with the best training point given by
the absolute minimum of the validation cost function $C_{\rm ZLP}^{(\rm val)}$ evaluated
over a sufficiently large number of iterations.

In order to estimate and propagate uncertainties associated
to the ZLP parametrisation and subtraction procedure,
here we adopt a variant of the Monte Carlo replica
method~\cite{Roest:2020kqy} benefiting from the high statistics
(large number of pixels) provided by an EELS-SI.
The starting point is selecting  $N_{\rm rep}\simeq \mathcal{O}\lp 5000\rp$
subsets of spectra such as the one in Eq.~(\ref{eq:random_choice_spectra_app})
containing one representative of each of the $K$ clusters considered.
One denotes this subset of spectra as a Monte Carlo (MC) replica,
and we denote the collection of replicas by
\be
\label{eq:random_choice_spectra_v2}
      {\boldsymbol I}^{(m)} = \left\{ I_{\rm EELS}^{(i_{m,1},j_{m,1})}(E), I_{\rm EELS}^{(i_{m,2},j_{m,2})}(E),\ldots, I_{\rm EELS}^{(i_{m,K},j_{m,K})}(E) \right\} \, ,
      \quad m=1,\ldots,N_{\rm rep} \, .
\ee
where now the superindices $(i_{m,k},j_{m,k})$ indicate a specific spectrum
from the $k$-th cluster that has been assigned to the $m$-th replica.
Given that these replicas are selected at random, they provide a representation
of the underlying probability density in the space of EELS spectra, e.g. those spectra
closer to the cluster mean will be represented more frequently in the replica
distribution.

    %

By training now a separate model 
to each of the $N_{\rm rep}$ replicas,
one ends up with another Monte Carlo representation, now 
of the probability density in the space of ZLP parametrisations.
This is done by replacing the cost function Eq.~(\ref{eq:costfunction_NNtraining_appendix}) by
\be
\label{eq:costfunction_NNtraining_v2}
C_{\rm ZLP}^{(m)} = \frac{1}{n_{E}}\sum_{k=1}^K \sum_{\ell_k=1}^{n_E^{(k)}} \frac{\lc
  I_{\rm EELS}^{(i_{m,k},j_{m,k})}(E_{\ell_k}) - I_{\rm ZLP}^{({\rm NN})(m)} \lp
  E_{\ell_k},\ln \lp N_{\rm tot}^{(i_{m,k},j_{m,k})}\rp \rp \rc^2}{\sigma^2_k \lp E_{\ell_k}\rp }  \, , \qquad
E_{\ell_k} \le E_{{\rm I},k} \, , 
\ee
and then performing the model training separately for each individual replica.
Note that the denominator of the cost function Eq.~(\ref{eq:costfunction_NNtraining_v2}) is
independent of the replica.
The resulting Monte Carlo distribution of ZLP models, indicated by
\be
\label{eq:random_choice_spectra_v6}
      {\boldsymbol I}_{\rm ZLP}^{(\rm NN)} = \left\{ I_{\rm ZLP}^{({\rm NN})(1)} \lp
      E,\ln \lp N_{\rm tot} \rp\rp, \ldots,
 I_{\rm ZLP}^{({\rm NN})(N_{\rm rep})} \lp
  E,\ln \lp N_{\rm tot} \rp\rp
      \right\} 
\ee
makes possible subtracting the ZLP from the measured EELS spectra
following the matching procedure described in~\cite{Roest:2020kqy}
and hence  isolating the inelastic contribution in each
pixel, 
\be
 I_{\rm inel}^{(i,j)(m)}(E) \simeq \lc I^{(i,j)}_{\rm EELS}(E) -  I_{\rm ZLP}^{({\rm NN})(m)} \lp
 E,\ln \lp N_{\rm tot}^{(i,j)} \rp \rp \rc \, ,\quad m=1,\ldots,N_{\rm rep}  \, .
\ee
The variance of $ I_{\rm inel}^{(i,j)}(E)$ over the MC replica sample
estimates the  uncertainties associated to the ZLP
subtraction procedure.
By means of these MC samplings of the  probability distributions associated
to the ZLP and inelastic components of the recorded spectra,
one can evaluate the relevant
derived quantities with a faithful error estimate.
Note that in our approach error propagation is realised without the need to resort to
any approximation, e.g. linear error analysis.

One important benefit of Eq.~(\ref{eq:costfunction_NNtraining_v2}) is that the machine learning
model training can be carried out fully in parallel, rather than sequentially, for each replica.
Hence our approach is most efficiently implemented when running on a computer cluster
with a large number of CPU (or GPU) nodes, since this configuration maximally
exploits the parallelization flexibility of the Monte Carlo replica method.

As mentioned above, the cluster-dependent hyperparameters $E_{{\rm I},k}$
ensure that the model is trained only in the  energy loss data region
where ZLP dominates total intensity.
This is illustrated by the scheme of Fig.~3.3 in~\cite{Roest:2020kqy}, which displays a toy simulation of the ZLP and inelastic scattering
contributions adding up to the total recorded EELS intensity.
The neural network
model for the ZLP is then trained on the data corresponding to region I,
while region II is obtained entirely from model predictions.
To determine the values of $E_{{\rm I},k}$, we evaluate the first derivative of the total
recording intensity, $dI_{\rm EELS}(E)/dE$, for each of the members
of the $k$-th cluster.
When this derivative crosses zero, the contribution from $I_{\rm inel}$ will
already be dominant.
There are then two options.
First, one sets $E_{{\rm I},k} = f \times E_{{\rm min},k}$, where $f < 1$
and $ E_{{\rm min},k}$ is the energy where the median of $dI_{\rm EELS}/dE$ crosses zero
(first local minimum) for cluster $k$.
Second, one sets 
  $E_{{\rm I},k} $ to be  the value where at most $f\%$ of the models
have crossed $dI_{\rm EELS}/dE=0$, with $f\simeq 10\%$.
This choice implies that 90\% of the models still exhibit a negative derivative.
We have verified that compatible results are obtained
with the two choices, indicating that results are reasonably stable
with respect to the value of the hyperparameter $E_{{\rm I},k}$.

The second model hyperparameter, denoted by $E_{{\rm II},k}$ in Fig.~3.3 in~\cite{Roest:2020kqy},
indicates the region for which the ZLP can be considered as fully negligible.
Hence in this region III we impose that $I_{\rm ZLP}(E)\to 0$ by means of the Lagrange multiplier
method.
This condition fixes the model behaviour in the large energy loss limit,
which otherwise would remain unconstrained.
Since the ZLP is known to be a steeply-falling function, $E_{{\rm II},k}$ should not chosen
not too far from $E_{{\rm I},k}$ to avoid an excessive interpolation region.
In this work we use $E_{{\rm II},k}=3\times E_{{\rm I},k}$, though this choice can be adjusted
by the user.

Finally, we mention that the model hyperparameters $E_{{\rm I},k}$ and
$E_{{\rm II},k}$ could eventually be determined by means of an automated
hyper-optimisation procedure as proposed in~\cite{Ball:2021leu}, hence
further reducing the need for human-specific input in the whole procedure.

\clearpage

\section{Kramers-Kronig analysis of EEL spectra}
\label{sec:theory}

Here we provide an overview of the 
theoretical formalism, based on~\cite{Egerton:2007}, adopted to evaluate
the single-scattering
distribution, local thickness, bandgap energy and type
and complex dielectric function from the measured EELS spectra.
As indicated by Eq.~(1) in the main manuscript,
these spectra  
receive three contributions:
the one from inelastic single scatterings off the electrons
in the specimen, the one  associated to multiple inelastic scatterings,
and then the ZLP arising from elastic scatterings and instrumental broadening.
Hence a generic EEL spectrum $I_{\rm EELS}(E)$
can be decomposed as
\begin{equation}
  \label{eq:EELSmaster}
  I_{\rm EELS}(E) =
I_{\rm ZLP}(E) +
  I_{\rm inel}(E) =
  I_{\rm ZLP}(E) +
  \sum_{n=1}^\infty I^{(n)}_{\rm inel}(E) \, ,
\end{equation}
where $E$ is the energy loss experienced by the electrons
upon traversing the specimen, $I_{\rm ZLP}$ is the ZLP
intensity, and $ I^{(n)}_{\rm inel}$ indicates the contribution associated
to $n$ inelastic scatterings.
The ZLP intensity can be further expressed as
\be
\label{eq:ZLP_norm}
I_{\rm ZLP}(E) = N_0 R(E) \, ,\qquad \int_{-\infty}^\infty dE \,R(E)=1 \, ,
\ee
where $R(E)$ is known as the resolution or instrumental response function
whose full width at half-maximum (FWHM) indicates the resolution of the instrument.
The normalisation factor $N_0$ thus corresponds to the integrated intensity under the zero-loss
peak.
In the following, we assume that the ZLP contribution to Eq.~(\ref{eq:EELSmaster}) has already
been disentangled from that associated to inelastic scatterings by means of the subtraction procedure
described in App.~\ref{app:nntraining}.

\paragraph{The single-scattering distribution.}
If one denotes by $t$ the local thickness of the specimen and by
$\lambda$ the mean free path of the electrons, then assuming that
inelastic scatterings are uncorrelated and that  $t \gsim \lambda$, one has that 
the integral over the $n$-scatterings distribution
$I^{(n)}_{\rm inel}$ is a Poisson distribution
\be
\label{eq:N_n}
N_n \equiv \int_{-\infty}^{\infty} dE\,
I^{(n)}_{\rm inel}(E) =
B \frac{\lp t/\lambda \rp^n}{n!}e^{-t/\lambda} \, , \qquad n=1,2,\ldots,\infty \, ,
\ee
with $B$ a normalisation constant.
From the combination of Eqns.~(\ref{eq:EELSmaster})
and~(\ref{eq:N_n}) it follows that
\be
\label{eq:norm_inelastic}
N_{\rm inel}\equiv \int_{-\infty}^{\infty} dE\,
I_{\rm inel}(E) = \sum_{n=1}^\infty N_n = B \sum_{n=1}^\infty
 \frac{\lp t/\lambda \rp^n}{n!}e^{-t/\lambda}=B\lp 1-e^{-t/\lambda}\rp \, ,
 \ee
 and hence one finds that the
 integral over the $n$-scatterings distribution is
 such that
 \be
N_n = \frac{N_{\rm inel}}{\lp 1-e^{-t/\lambda} \rp}\frac{\lp t/\lambda \rp^n}{n!}e^{-t/\lambda} \, ,
\ee
in terms of the normalisation $N_{\rm inel}$ of the full inelastic
scattering distribution, the sample thickness $t$
and the mean free path $\lambda$.
Note also that the ZLP normalisation factor $N_0$ is then given in terms of the inelastic one as
\be
N_0 = \frac{N_{\rm inel}}{e^{t/\lambda}-1} \, ,
\ee
and hence one has the following relations between integrated
inelastic scattering intensities
\be
\frac{N_1^n}{N_n}=n!N_0^{n-1} \, ,\qquad \forall ~n \ge 1 \, .
\ee

In order to evaluate the local thickness of the specimen
and the corresponding dielectric function, it is necessary
to deconvolute the measured spectra and extract from them
the single-scattering distribution (SSD), $I_{\rm SSD}(E)$.
The SSD is related to the experimentally measured $n=1$ 
distribution, $I^{(1)}_{\rm inel}(E)$
by the finite resolution of our measurement apparatus:
\be
\label{eq:def_convolution}
I^{(1)}_{\rm inel}(E) = R(E)\otimes I_{\rm SSD}(E) \equiv
\int_{-\infty}^{\infty} dE'\, R(E-E')I_{\rm SSD}(E') \, ,
\ee
where in the following $\otimes$ denotes the convolution operation.
It can be shown, again treating individual scatterings as uncorrelated,
that the experimentally measured $n=2$ and $n=3$ multiple scattering distributions
can be expressed in terms of the SSD
as
\bea
I^{(2)}_{\rm inel}(E) &=&  R(E)\otimes I_{\rm SSD}(E)\otimes I_{\rm SSD}(E)/\lp 2! N_0\rp \ ,
\\
I^{(3)}_{\rm inel}(E) &=&
R(E)\otimes I_{\rm SSD}(E)\otimes I_{\rm SSD}(E)\otimes I_{\rm SSD}(E)/\lp 3! N^2_0\rp \ ,
\eea
and likewise for $n\ge 4$.
Combining this information, one observes that the
spectrum Eq.~(\ref{eq:EELSmaster}) can be expressed in terms of the resolution
function $R$, the ZLP normalisation $N_0$,
and the single-scattering distribution $I_{\rm SSD}$ as follows
\bea
  \label{eq:EELSmaster_2}
 && I_{\rm EELS}(E) \nonumber
  = N_0 R(E) + R(E)\otimes I_{\rm SSD}(E) + R(E)\otimes I_{\rm SSD}(E)\otimes I_{\rm SSD}(E)/\lp 2! N_0\rp + \ldots\\ \nonumber
  && =R(E) \otimes \lp N_0\delta(E) + I_{\rm SSD}(E) + I_{\rm SSD}(E)\otimes I_{\rm SSD}(E)/\lp 2! N_0\rp +\ldots \rp \\
 && =N_0 R(E) \otimes \lp \delta(E) +\sum_{n=1}^{\infty} \lc I_{\rm SSD}(E)\otimes\rc^n \delta(E)/\lp n! N_0^{n}\rp  \rp \, ,
\eea
where $\delta(E)$ is the Dirac delta function.
If the ZLP normalisation factor $N_0$ and resolution function $R(E)$ are known, then one can use
Eq.~(\ref{eq:EELSmaster_2}) to extract the SSD from the measured
spectra by means of a deconvolution procedure.

\paragraph{SSD deconvolution.}
The structure of Eq.~(\ref{eq:EELSmaster_2}) indicates that transforming to Fourier space
will lead to an algebraic equation which can then be solved for the SSD.
Here we define the Fourier transform $\widetilde{f}(\nu)$ of a function $f(E)$ as follows
\be
\label{eq:continuous_fourier_transform}
\mathcal{F}\lc f(E) \rc(\nu)\equiv \widetilde{f}(\nu)\equiv \int_{-\infty}^\infty dE\,f(E) e^{-2\pi i E\nu}\, ,
\ee
whose inverse is given by
\be
\label{eq:continuous_fourier_transform_inverse}
\mathcal{F}^{-1}\lc \widetilde{f}(\nu) \rc(E)=
f(E)\equiv \int_{-\infty}^\infty d\nu\,\widetilde{f}(\nu) e^{2\pi i E\nu}\, ,
\ee
which has the useful property that convolutions such as  Eq.~(\ref{eq:def_convolution})
are transformed into products,
\be
\label{eq:fourier_convolutions}
{\rm if~}f(E)=g(E)\otimes h(E)\quad{\rm then}\quad \mathcal{F}\lc f(E) \rc = \widetilde{f}(\nu) = \widetilde{g}(\nu)\widetilde{h}(\nu) \, .
\ee
The Fourier transform of Eq.~(\ref{eq:EELSmaster_2}) leads to
the Taylor expansion of the exponential and hence
\be
\widetilde{I}_{\rm EELS}(\nu)=N_0\widetilde{R}(\nu)\exp\lp  \frac{\widetilde{I}_{\rm SSD}(\nu)}{N_0}\rp \, ,
\ee
which can be solved for the Fourier transform of the single scattering distribution
\be
\widetilde{I}_{\rm SSD}(\nu)=N_0 \ln \frac{\widetilde{I}_{\rm EELS}(\nu)}{N_0\widetilde{R}(\nu)}
= N_0 \ln \frac{\mathcal{F}\lc I_{\rm EELS}(E)\rc (\nu)}{N_0 \mathcal{F}\lc R(E)\rc (\nu)  } \, .
\ee
By taking the  inverse Fourier transform, one obtains the sought-for expression
for the single scattering distribution as a function of the electron energy loss
\be
\label{eq:deconvolution_procedure}
I_{\rm SSD}(E)=N_0 \mathcal{F}^{-1}\lc \ln \frac{\mathcal{F}\lc I_{\rm EELS}\rc }{N_0 \mathcal{F}\lc  R\rc
}\rc \, ,
\ee
where the only required inputs are the experimentally measured EELS spectra,
Eq.~(\ref{eq:EELSmaster}), with the corresponding ZLP.

\paragraph{Discrete Fourier transforms.} 
Eq.~(\ref{eq:deconvolution_procedure}) can be evaluated
numerically by approximating the continuous transform
Eq.~(\ref{eq:continuous_fourier_transform}) by its discrete Fourier transform
equivalent.
The discrete Fourier transform  of a discretised function $f(E)$
defined at $E_n \in \{E_0, ..., E_{N-1}\}$  is given by:
\begin{equation}
  \label{eq_def_DFT}
\mathcal{F}_D \lc f(E) \rc (\nu_k) = \widetilde{f}(\nu_k) = \sum^{N-1}_{n=0} \operatorname{e}^{-i2\pi kn/N} f(E_n), \qquad \forall\, k \in \{0, ..., N-1\}\,,
\end{equation}
with the corresponding inverse transformation being
\begin{equation}
\label{eq_def_DFT_inverse}
\mathcal{F}_D^{-1} \lc \widetilde{f}(\nu) \rc (E_n) = f(E_n) =\frac{1}{N} \sum^{N-1}_{k=0} \operatorname{e}^{i2\pi kn/N}  \widetilde{f}(\nu_k) \qquad \forall\, n \in \{0, ..., N-1\}\,.
\end{equation}
If one approximates the continuous function $f(E)$ by its discretised version
$f(E_0 + n\Delta E)$ and likewise $\widetilde{f}(\nu)$ by $\widetilde{f}(k\Delta \nu)$ where
$\Delta x \Delta \nu = N^{-1}$ one finds that
\begin{equation}\label{eq_approx_CFT}
	\widetilde{f}(\nu) \approx \Delta x e^{-i 2\pi k \Delta \nu E_0}\mathcal{F}_D \lc f(E)\rc \,,
\end{equation}
and likewise for the inverse transform
\be
f(E) \approx \frac{1}{\Delta x}
\mathcal{F}_D^{-1} \lc \widetilde{g}(k\Delta\nu) \rc \, ,\qquad
\widetilde{g}(k\Delta\nu) \equiv
e^{i2\pi k \Delta\nu E_0} \widetilde{f}(k\Delta\nu) \, .
\ee
In practice, the EELS spectra considered are characterised by a fine
spacing in $E$ and the discrete approximation for the Fourier
transform produces
results very close to the exact one.


\paragraph{Thickness calculation.}
\label{subsec:thickness_calculation}
Once the SSD has been determined by means of the deconvolution procedure
summarised by Eq.~(\ref{eq:deconvolution_procedure}), it can be used
as input in order to evaluate the local sample thickness $t$ from the experimentally measured
spectra.
Kramers-Kronig analysis  provides the following relation between the thickness $t$,
the ZLP normalisation $N_0$, and the single-scattering
distribution,
\be
\label{eq:thickness_calculation}
t = \frac{4a_0 F E_0}{N_0\lp  1-{\rm Re}\lc 1/\epsilon(0)\rc\rp}
\int_0^\infty dE\frac{I_{\rm SSD}(E)}{E\ln \lp 1+\beta^2/\theta_E^2\rp} \, ,
\ee
where we have assumed that the effects of surface scatterings can be neglected.
In Eq.~(\ref{eq:thickness_calculation}), $a_0=0.0529$ nm is Bohr's radius, $F$ is a relativistic
correction factor,
\be
F = \frac{  1+E_0/(1022~{\rm keV})  }{\lc 1+E_0/(511~{\rm keV})\rc^2  } \, ,
\ee
with $E_0$ being the incident electron energy, $\epsilon(E)$ is the complex dielectric function,
and $\theta_E$ is the characteristic angle defined by
\be
\label{eq:characteristic_angle}
\theta_E = \frac{E}{\gamma m_0v^2} = \frac{E}{\lp E_0 + m_0c^2\rp (v/c)^2}
\ee
with $\gamma$ being the usual relativistic dilation factor, $\gamma=\lp 1-v^2/c^2\rp^{-1/2}$,
and $\beta$ the collection semi-angle of the microscope.\footnote{Which should
not be confused with the normalised velocity often used in relativity, $\beta=v/c$.}
For either an  insulator or a semiconductor material
with refractive index $n$, one has that
\be
\label{eq:refractive_index}
{\rm Re}\lc 1/\epsilon(0)\rc = n^{-2} \, ,
\ee
while ${\rm Re}\lc 1/\epsilon(0)\rc=0$ for a metal or semi-metal.
Hence, the determination of the dielectric function is not a pre-requisite to evaluate the
specimen thickness,
and for given microscope operation conditions we can express Eq.~(\ref{eq:thickness_calculation})
as
\be
\label{eq:thickness_calculation_v2}
t = \frac{A}{N_0}
\int_0^\infty dE\frac{I_{\rm SSD}(E)}{E\ln \lp 1+\beta^2/\theta_E^2\rp} \, ,
\ee
with $A$  constant across the specimen.
If the thickness of the specimen
is already known at some location, then Eq.~(\ref{eq:thickness_calculation_v2})  can be  used
to calibrate $A$ and  evaluate this thickness elsewhere.
Furthermore, if the thickness of the material has already been
determined by means of an independent experimental technique, then
Eq.~(\ref{eq:thickness_calculation}) can be inverted to determine the refractive index $n$
of an insulator or semi-conducting material using
\be
n = \lc 1-\frac{4a_0 FE_0}{N_0 t} \lp \int_0^\infty dE\frac{I_{\rm SSD}(E)}{E\ln \lp 1+\beta^2/\theta_E^2\rp} \rp \rc^{-1/2} \, .
\ee


\paragraph{The complex dielectric function.}
\label{sec:KKanalysis}
The
dielectric function of a material, also known as permittivity, is a measure of how
easy or difficult it is to polarise a dielectric material such an insulator upon the
application of an external electric field.
In the case of oscillating electric fields such as those that constitute electromagnetic radiation,
the dielectric response will have both a real and a complex part and will depend
on the oscillation frequency $\omega$,
\be
\epsilon(\omega)={\rm Re}\lc \epsilon(\omega)\rc+i{\rm Im}\lc \epsilon(\omega)\rc \, ,
\ee
which can also be expressed in terms of the energy $E=\hbar \omega$ of the photons
that constitute this electromagnetic radiation,
\be
\label{eq:dielectric_function_def}
\epsilon(E)={\rm Re}\lc \epsilon(E)\rc+i{\rm Im}\lc \epsilon(E)\rc \, .
\ee
In the vacuum, the real and imaginary parts
of the dielectric function reduce to ${\rm Re}\lc \epsilon(E)\rc=1$
and ${\rm Im}\lc \epsilon(E)\rc=0$.
Furthermore, the dielectric function is related to the susceptibility $\chi$ by
\be
\epsilon(E)=1-\nu\chi(E) \, , 
\ee
where $\nu$ is the so-called Coulomb matrix.

The single scattering distribution $I_{\rm SSD}(E)$ is related to the imaginary
part of the  complex dielectric function $\epsilon(E)$ by means the following relation
\be
I_{\rm SSD}(E) = \frac{N_0 t}{\pi a_0 m_0 v^2}{\rm Im}\lc \frac{-1}{\epsilon(E)}\rc
\ln \lc 1+\lp \frac{\beta}{\theta_E}\rp^2\rc \, ,
\ee
in terms of the sample thickness $t$, the ZLP normalisation $N_0$, and
the microscope operation parameters defined in
Sect.~\ref{subsec:thickness_calculation}.
We can invert this relation to obtain
\be
\label{eq:im_diel_fun}
{\rm Im}\lc \frac{-1}{\epsilon(E)}\rc = \frac{\pi a_0 m_0 v^2}{N_0 t}\frac{I_{\rm SSD}(E)}{\ln \lc 1+\lp \frac{\beta}{\theta_E}\rp^2\rc} \, .
\ee
Since the prefactor in Eq.~(\ref{eq:im_diel_fun}) does not depend on the energy loss $E$,
we see that ${\rm Im}[-1/\epsilon(E)]$ will be proportional to the single scattering
distribution $I_{\rm SSD}(E)$ with a denominator that decreases with the energy
(since $\theta_E\propto E$)
and hence weights more higher energy losses.

Given that the dielectric response function is causal, the real part of the dielectric function
can be obtained from the imaginary one by using a Kramers-Kronig relation of the form
\be
\label{eq:kramerskronig}
{\rm Re}\lc \frac{1}{\epsilon(E)}\rc = 1-\frac{2}{\pi}\mathcal{P}\int_0^{\infty}  dE'\, {\rm Im}
\lc \frac{-1}{\epsilon(E')}\rc \frac{E'}{E'^2-E^2} \, ,
\ee
where $\mathcal{P}$ stands for Cauchy's prescription to evaluate the principal
part of the integral.
A particularly important application of this relation is the $E=0$ case,
\be
\label{eq:normalisation_im_deltaEim}
{\rm Re}\lc \frac{1}{\epsilon(0)}\rc = 1-\frac{2}{\pi}\mathcal{P}\int_0^{\infty}  dE\, {\rm Im}
\lc \frac{-1}{\epsilon(E)}\rc \frac{1}{E} \, ,
\ee
which is known as the Kramers-Kronig sum rule.
Eq.~(\ref{eq:normalisation_im_deltaEim}) can be used to determine the overall
normalisation of $ {\rm Im}\lc -1/\epsilon(E)\rc$, since
${\rm Re}\lc 1/\epsilon(0)\rc$ is known for most materials.
For instance, as mentioned in Eq.~(\ref{eq:refractive_index}), for an insulator
or semiconductor material it is given in terms of its refractive index $n$.

Once the imaginary part of the dielectric function has been determined
from the single-scattering distribution, Eq.~(\ref{eq:im_diel_fun}),
then one can obtain the corresponding real part by means of the Kramers-Kronig relation,
Eq.~(\ref{eq:kramerskronig}).
Afterwards, the full complex dielectric function can be reconstructed
by combining the calculation of the real and imaginary parts, since
\be
\epsilon(E)={\rm Re}\lc \epsilon(E)\rc+i{\rm Im}\lc \epsilon(E)\rc \equiv
\epsilon_1(E)+i\epsilon_2(E) \, ,
\ee
implies that
\be
\label{eq:dielectric_real_imaginary}
{\rm Re}\lc \frac{1}{\epsilon(E)}\rc = \frac{\epsilon_1(E)}{\epsilon_1^2(E) + \epsilon_2^2(E)}\,,\qquad
{\rm Im}\lc \frac{-1}{\epsilon(E)}\rc = \frac{\epsilon_2(E)}{\epsilon_1^2(E) + \epsilon_2^2(E)}\,,
\ee
and hence one can express the dielectric function in terms of
experimentally accessible quantities,
\be
\label{eq:final_dielectric_function}
\epsilon(E) = \frac{{\rm Re}\lc \frac{1}{\epsilon(E)}\rc+ i{\rm Im}\lc \frac{-1}{\epsilon(E)}\rc}{\lp {\rm Re}\lc \frac{1}{\epsilon(E)}\rc\rp^2+\lp {\rm Im}\lc \frac{-1}{\epsilon(E)}\rc\rp^2} \, .
\ee
Once the complex dielectric function of a material has been determined, it is possible
to evaluate related quantities that also provide information about the opto-electronic
properties of a material.
One example of this would be the optical absorption coefficient, given by
\be
\mu(E) = \frac{E}{\hbar c}\lc 2\lp \epsilon_1^2(E)+\epsilon_2^2(E)\rp^{1/2}-2\epsilon_1(E)\rc^{1/2} \,,
\ee
which represents a measure of how far light of a given wavelength $\lambda=hc/E$ can penetrate
into a material before it is fully extinguished via absorption processes.
Furthermore, combining Eqns.~(\ref{eq:refractive_index}) and~(\ref{eq:dielectric_real_imaginary})
one has that for a semiconductor material, such as those considered in this work, the refractive
index is given by the relation
\be
\label{eq:refractiveindex_calculation}
n = \lp  \frac{\epsilon_1(0)}{\epsilon_1^2(0) + \epsilon_2^2(0)} \rp^{-1/2} \, ,
\ee
which implies a positive, non-zero value of the real part of the complex dielectric function
at $E=0$.

The complex dielectric function  $\epsilon(E)$
provides direct information on the opto-electronic properties
of a material, for example those associated to plasmonic resonances.
Specifically, a collective plasmonic excitation should be indicated by the condition
that the real part of the dielectric function crosses the $x$ axis, $\epsilon_1(E)=0$,
with a positive slope.
These plasmonic excitations typically are also translated by a well-defined peak
in the energy loss spectra.
Hence, verifying that a plasmonic transition indicated by  $\epsilon_1(E)=0$
corresponds to specific energy-loss features
provides a valuable handle to pinpoint the nature of local electronic excitations
present in the analysed specimen.

\paragraph{The role of surface scatterings.}
The previous derivations assume that the specimen is thick enough such that
the bulk of the measured energy loss distributions arises from volume
inelastic scatterings, while edge- and surface-specific contributions
can be neglected.
However, for relatively thin samples with thickness $t$ below a few tens of nm,
this approximation is not necessarily suitable.
Assuming a locally flat specimen with two surfaces,
in this case  Eq.~(\ref{eq:EELSmaster}) must
be generalised to
\begin{equation}
  \label{eq:EELSmaster_v3}
  I_{\rm EELS}(E) =
I_{\rm ZLP}(E) +
  I_{\rm inel}(E) +  I_{S}(E) 
\end{equation}
with  $ I_{S}(E)$ representing the  contribution from surface-specific
inelastic scattering.
This surface contribution can be
evaluated in terms of the real $\epsilon_1$ and imaginary $\epsilon_2$
components of the complex dielectric function,
\be
\label{eq:surface_intensity}
I_{S}(E) = \frac{N_0}{\pi a_0 k_0 T}\lc \frac{\tan^{-1}(\beta/\theta_E)}{\theta_E} -
\frac{\beta}{\beta^2+\theta_E^2}\rc \lp \frac{4\epsilon_2}{\lp \epsilon_1+1\rp^2
+\epsilon_2^2} - {\rm Im}\lc\frac{-1}{\epsilon(E)}\rc\rp \, ,
\ee
where the electron kinetic energy is $T=m_ev^2/2$.

The main challenge to evaluate the surface component from Eq.~(\ref{eq:surface_intensity}) is that
it depends on the complex dielectric function $\epsilon(E)$, which in turn is a function
of the single scattering distribution obtained from the deconvolution of $I_{\rm inel}(E) $
obtained assuming that $I_S(E)$ vanishes.
For not too thin specimens, the best approach is then an iterative procedure,
whereby one starts
by assuming that $I_{S}(E)\simeq 0$,
evaluates $\epsilon(E)$, and uses it to evaluate a first approximation to $I_S(E)$
using 
Eq.~(\ref{eq:surface_intensity}).
This approximation is then subtracted from Eq.~(\ref{eq:EELSmaster_v3})
and hence provides a better estimate of the bulk contribution $ I_{\rm inel}(E)$.
One can then
iterate the procedure until some convergence criterion is met.
Whether or not this procedure converges will depend on the specimen under consideration,
and specifically on the features of the EELS spectra at low energy losses,
$E\lsim 10$ eV.
For the specimens considered in this work, it is found that this iterative procedure
to determine the surface contributions converges best provided that the local sample
thickness satisfies $t \gsim 20$ nm.

\begin{figure}[ht]
\begin{centering}
  \includegraphics[width=0.49\linewidth]{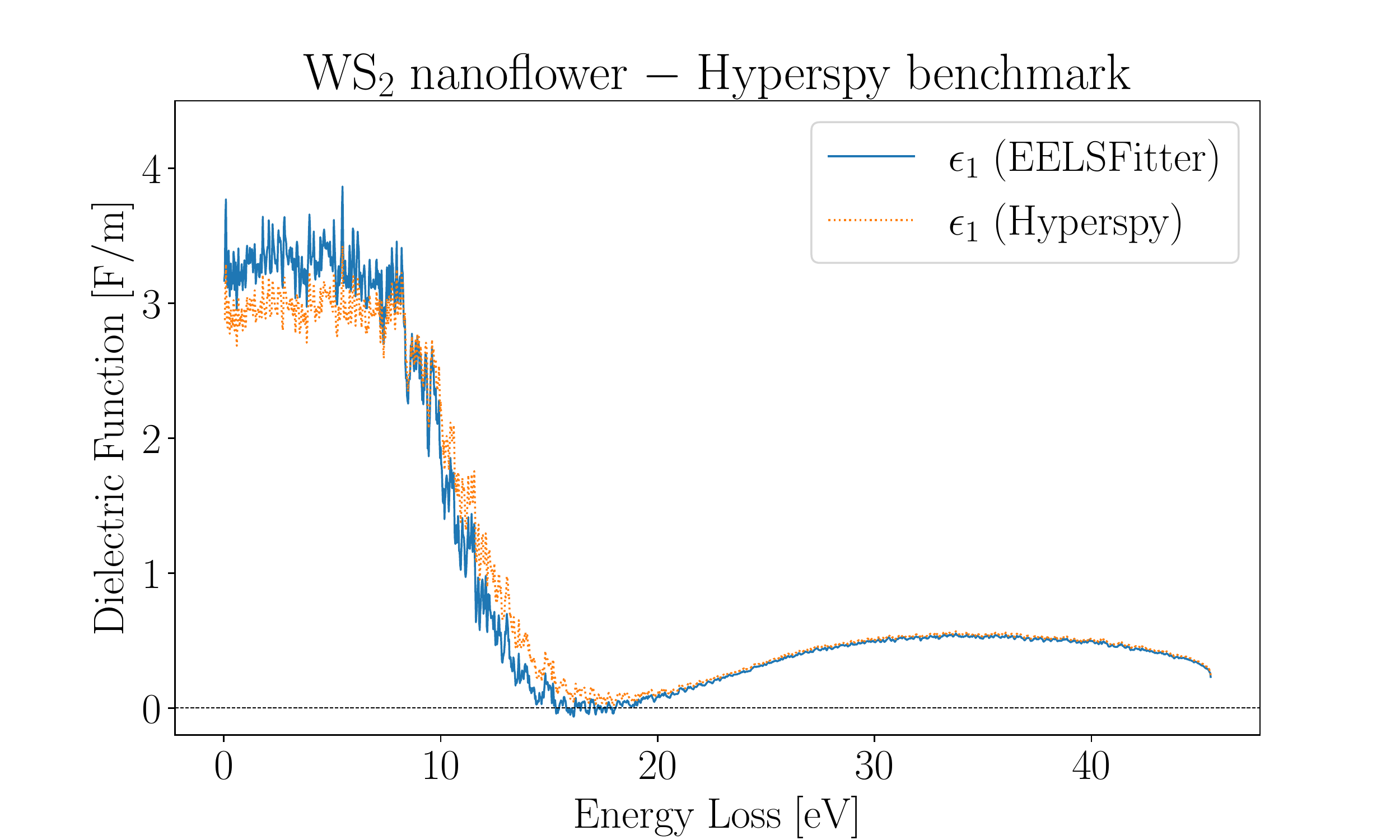}
  \includegraphics[width=0.49\linewidth]{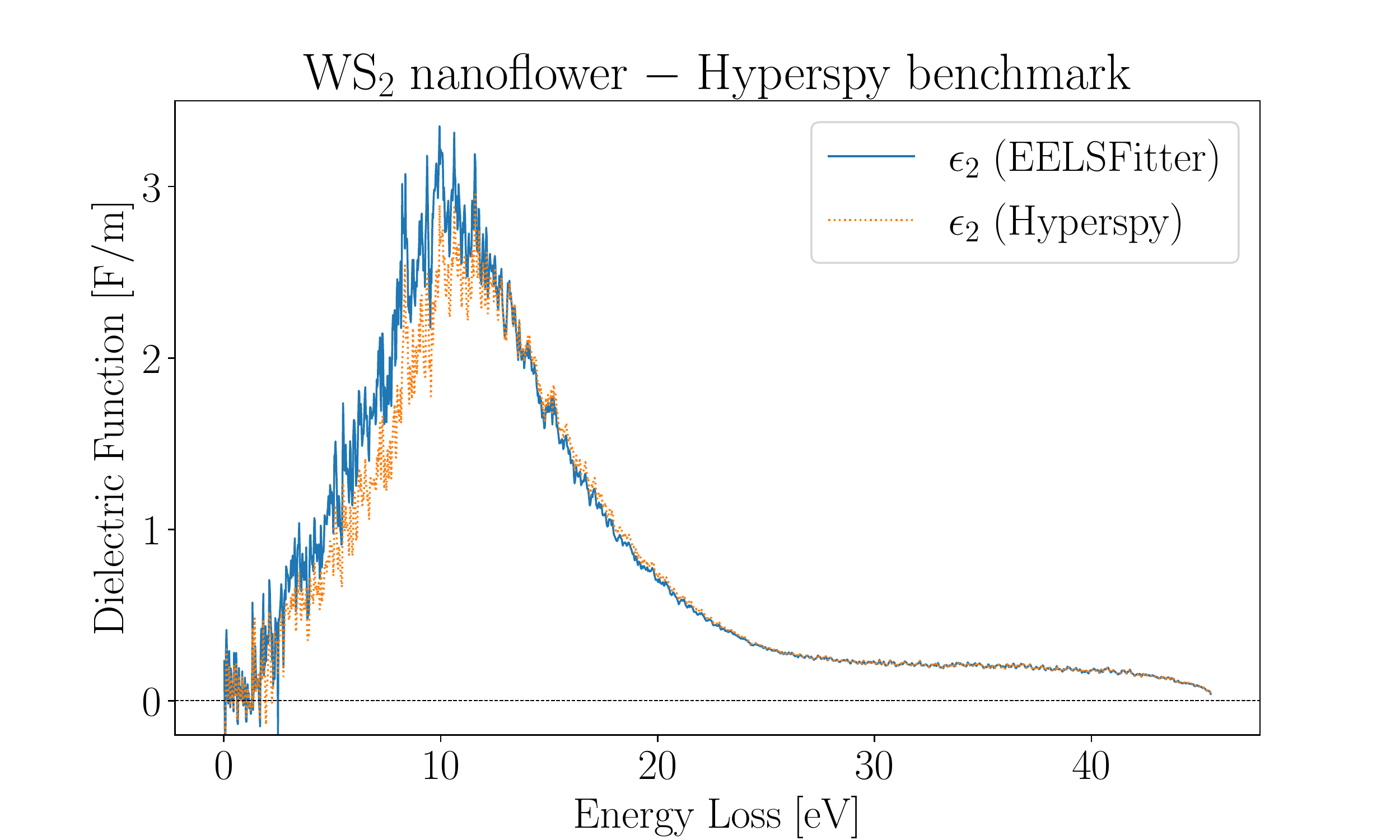}
  \caption{Comparison of the {\sc\small EELSfitter}-based results
for the real (left) and imaginary (right panel) components
of the dielectric function with those available from {\sc\small HyperSpy}
for the same input ZLP and inelastic spectrum, associated
to a representative pixel of the WS$_2$ nanoflower specimen.
  }
\label{fig:benchmark-hyperspy}
\end{centering}
\end{figure}

\paragraph{Validation.}
The calculations of the local specimen thickness, Eq.~(\ref{eq:thickness_calculation}), and of the complex dielectric function, Eq.~(\ref{eq:final_dielectric_function}) presented in this work have been benchmarked
with the corresponding implementation available within the {\sc\small HyperSpy} framework~\cite{hyperspy}.
Provided one inputs the same inelastic spectra and
ZLP parametrisation, 
agreement between the two calculations is obtained.
This benchmark is illustrated in
Fig.~\ref{fig:benchmark-hyperspy}, which
compares the {\sc\small EELSfitter}-based results
with those available from {\sc\small HyperSpy} separately
for the real  and imaginary components
of the dielectric function.
Both calculations use
for the same input ZLP and inelastic spectra, associated
to a representative pixel of the WS$_2$ nanoflower specimen.
Residual differences can be attributed to
implementation differences e.g. for the discrete
Fourier transforms.
This validation test further confirms the robustness
of the calculations presented in this work.

\clearpage

\section{Band gap analysis of the EELS low-loss region}
\label{sec:bandgap_analysis}

One important  application of ZLP-subtracted EELS spectra is
the determination of the bandgap energy and type (direct or indirect) in semiconductor
materials.
The reason is that the onset of the inelastic scattering intensity 
provides
information on the value of the bandgap energy $E_{\rm bg}$, while its shape
for $E\gsim E_{\rm bg}$ is determined by the underlying band structure.
 Different approaches  have been put forward to evaluate $E_{\rm bg}$ from 
subtracted EEL spectra, such as by means of the inflection point of the rising intensity or
a linear fit to the maximum positive slope~\cite{Schamm:2003},
see also~\cite{https://doi.org/10.1002/adfm.202105426}.
Following~\cite{Roest:2020kqy},
here we adopt the method of~\cite{Rafferty:1998,Rafferty:2000},
 where the behaviour
of $I_{\rm inel}(E)$ in the region close
to the onset of the inelastic scatterings is described by
\begin{equation}
  \label{eq:I1}
    I_{\rm inel}(E) \simeq  A \lp E-E_{\rm bg} \rp^{b} \, , \quad E \gsim E_{\rm bg} \, ,
\end{equation}
and vanishes for $E < E_{\rm bg}$.
Here $A$ is a normalisation constant, while the exponent $b$ provides
information on the type of bandgap: it is expected to be  $b\simeq0.5~(1.5)$ for a
semiconductor material characterised
by a direct~(indirect) bandgap.
While Eq.~(\ref{eq:I1}) requires as input the complete inelastic distribution,
in practice the onset region is dominated by the single-scattering distribution,
since  multiple scatterings
contribute only at higher energy losses.

The bandgap energy $E_{\rm bg}$, the overall normalisation factor $A$, and
the bandgap exponent $b$ can be determined from a least-squares fit to the
experimental data
on the ZLP-subtracted spectra.
This polynomial fit is carried out in the energy loss region around the bandgap
energy, $ [ E^{(\rm fit)}_{\rm min}, E^{(\rm fit)}_{\rm max}]$.
A judicious choice of this interval is necessary to achieve stable results:
a too wide energy range will bias the fit by probing regions
where Eq.~(\ref{eq:I1}) is not necessarily valid,
while a too narrow fit range might not contain sufficient information to stabilize the results
and be dominated by statistical fluctuation.

\begin{figure}[ht]
    \centering
    \includegraphics[width=0.95\textwidth]{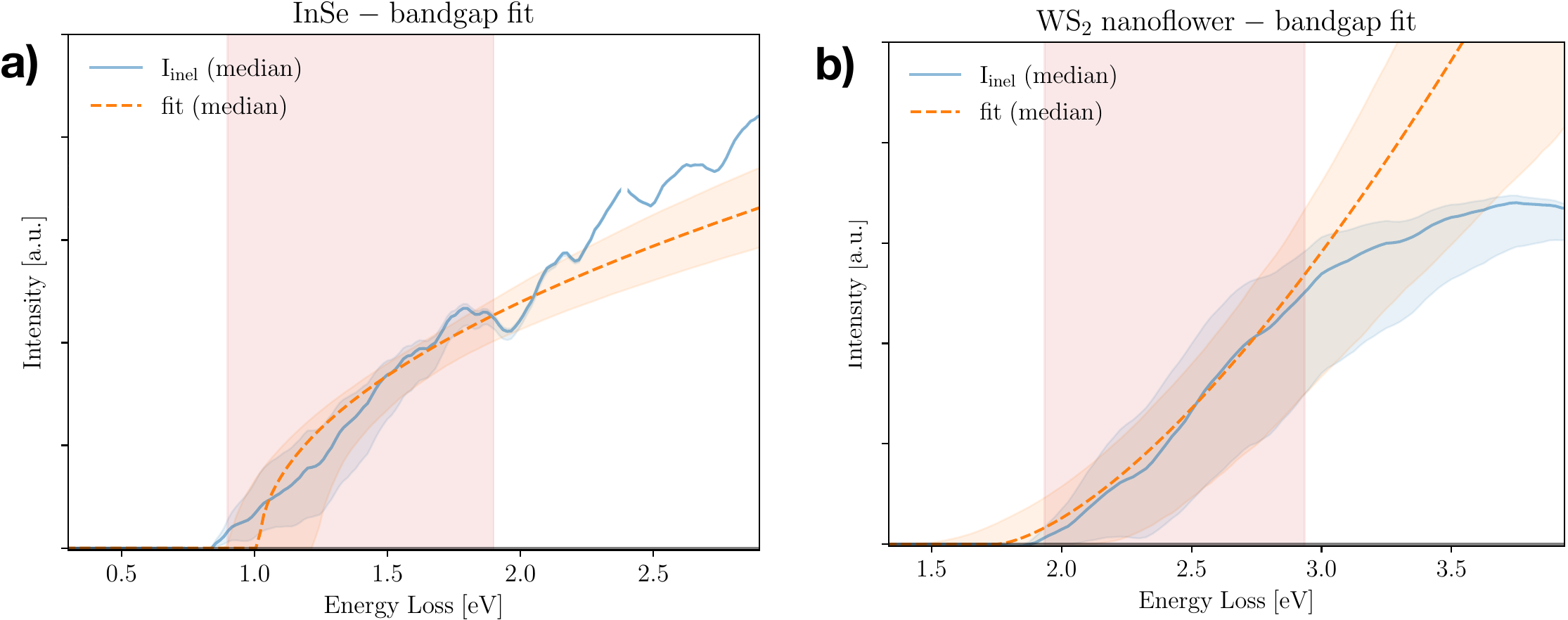}
    \caption{Representative examples of bandgap fits to the onset
      of inelastic spectra in the InSe (a) and WS$_2$ (b) specimens.
      The red shaded areas indicate the polynomial fitting range, the blue
      curve and band corresponds to
      the median and 68\% CL intervals of the 
       ZLP-subtracted intensity $I_{\rm inel}(E)$, and the outcome 
      of the bandgap fits based on Eq.~(\ref{eq:I1}) is
      indicated by the orange dashed curve (median) and band (68\% CL intervals).
    }
    \label{fig:Fig-SI5}
\end{figure}

Fig.~\ref{fig:Fig-SI5}(a,b) displays representative examples of bandgap fits to the onset
of the inelastic spectra in the InSe  and WS$_2$ specimens respectively.
The red shaded areas indicate the fitting range, bracketed
by  $E^{(\rm fit)}_{\rm min}$ and $E^{(\rm fit)}_{\rm max}$.
The blue
curve and band corresponds to
the median and 68\% CL intervals of the 
ZLP-subtracted intensity $I_{\rm inel}(E)$, and the outcome 
of the bandgap fits based on Eq.~(\ref{eq:I1}) is
indicated by the green dashed curve (median) and band (68\% CL intervals).
Here the onset exponents $b$ have been kept fixed to $b=0.5~(1.5)$ for the InSe~(WS$_2$)
specimen given the direct (indirect) nature of the underlying band-gaps.
One observes how the fitted model describes well the behaviour of $I_{\rm inel}(E)$
in the onset region for both specimens, further confirming the reliability of
our strategy to determine the bandgap energy $E_{\rm bg}$.
As mentioned in~\cite{Rafferty:2000}, it is important to avoid
taking a too large interval for  $ [ E^{(\rm fit)}_{\rm min}, E^{(\rm fit)}_{\rm max}]$,
else the polynomial
approximation ceases to be valid,  as one can also see directly
from these plots.

\clearpage

\section{Structural characterisation of the InSe specimen}
\label{sub:app-inse-hrtem}

Here we provide details on the structural characterisation
of the $n$-doped InSe specimens.
Each specimen is composed by a InSe nanosheet exhibiting a range of thicknesses.
The electronic properties of InSe, such as the  band gap value and
type, are sensitive to both the layer stacking
($\beta$, $\gamma$, or $\varepsilon$-phase)
as well as to the magnitude and type of doping~\cite{GURBULAK2014106,JULIEN2003263,Rigoult:a18761,doi:10.1021/nn405036u}.
In particular, $n$-doped $\varepsilon$-phase InSe has been
reported to exhibit a direct bandgap with value
$E_{\rm bg}=1.25$ eV~\cite{henck2019evidence}.

These InSe specimens have been
grown by means of the Bridgman-Stockbarger method.
Doping with Sn impurities is used to obtain $n$-type InSe.
InSe flakes are obtained from bulk material by the sonication
procedure~\cite{https://doi.org/10.1002/smll.201800749},
whereby single InSe crystals are pulverized and added to IPA with a ratio of 2:1 (mg:ml).
This combination is then sonicated in a sonic bath for 6 hours while keeping
the temperature in the range
between 25 $^\circ$C and 35 $^\circ$C.
The ultra high frequencies lead to gas formation between the layers of the material,
building up pressure
until adjacent layers split apart. 
The flakes in the resulting suspension are then collected and
dispersed on top of a TEM grid by pipetting.

\begin{figure}[t]
\begin{centering}
  \includegraphics[width=0.80\linewidth]{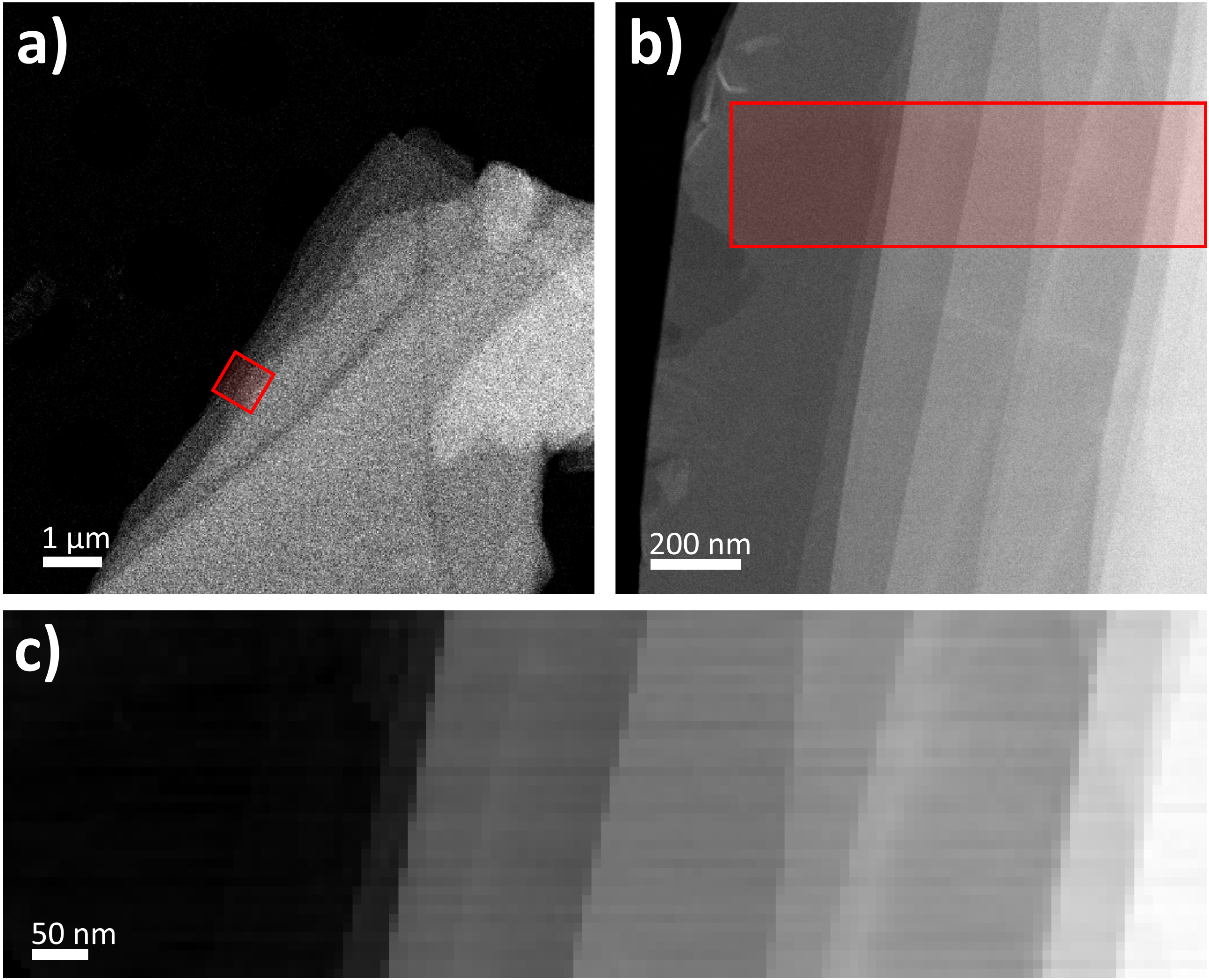}
  \vspace{0.2cm}
  \caption{(a) Low-magnification HAADF-STEM image of one of the $n$-doped InSe
    specimens analysed in this work.
    (b) Magnification of the region indicated with a red square in (a).
    (c) EELS-SI acquired on the region indicated with a red rectangle
    in (b).
    Note that the grey-scale is different in each  panel.
  }
\label{fig:FigSI-3}
\end{centering}
\end{figure}

Fig.~\ref{fig:FigSI-3}(a) displays a low-magnification 
High-Angle Annular Dark Field (HAADF) STEM image of one of the 
specimens obtained from this procedure.
The flake is lying on top of the holey carbon grid, and most of its volume is on top
of the vacuum.
Fig.~\ref{fig:FigSI-3}(b) shows then a
magnification of the region indicated with a red square in (a), while
in turn the red rectangle in (b) marks the region
where the corresponding EELS-SI, provided in Fig.~\ref{fig:FigSI-3}(c),
has been extracted.
The spatial resolution in this EELS-SI is around 8 nm.
Note that the (artificial) grey-scale convention adopted
is different in Figs.~\ref{fig:FigSI-3}(b) and~(c).
It can be observed that most of the flake
turns out to be bulk, exhibiting thicknesses of several monolayers at least,
with some thinner regions at the edges.

\begin{figure}[t]
\begin{centering}
  \includegraphics[width=0.80\linewidth]{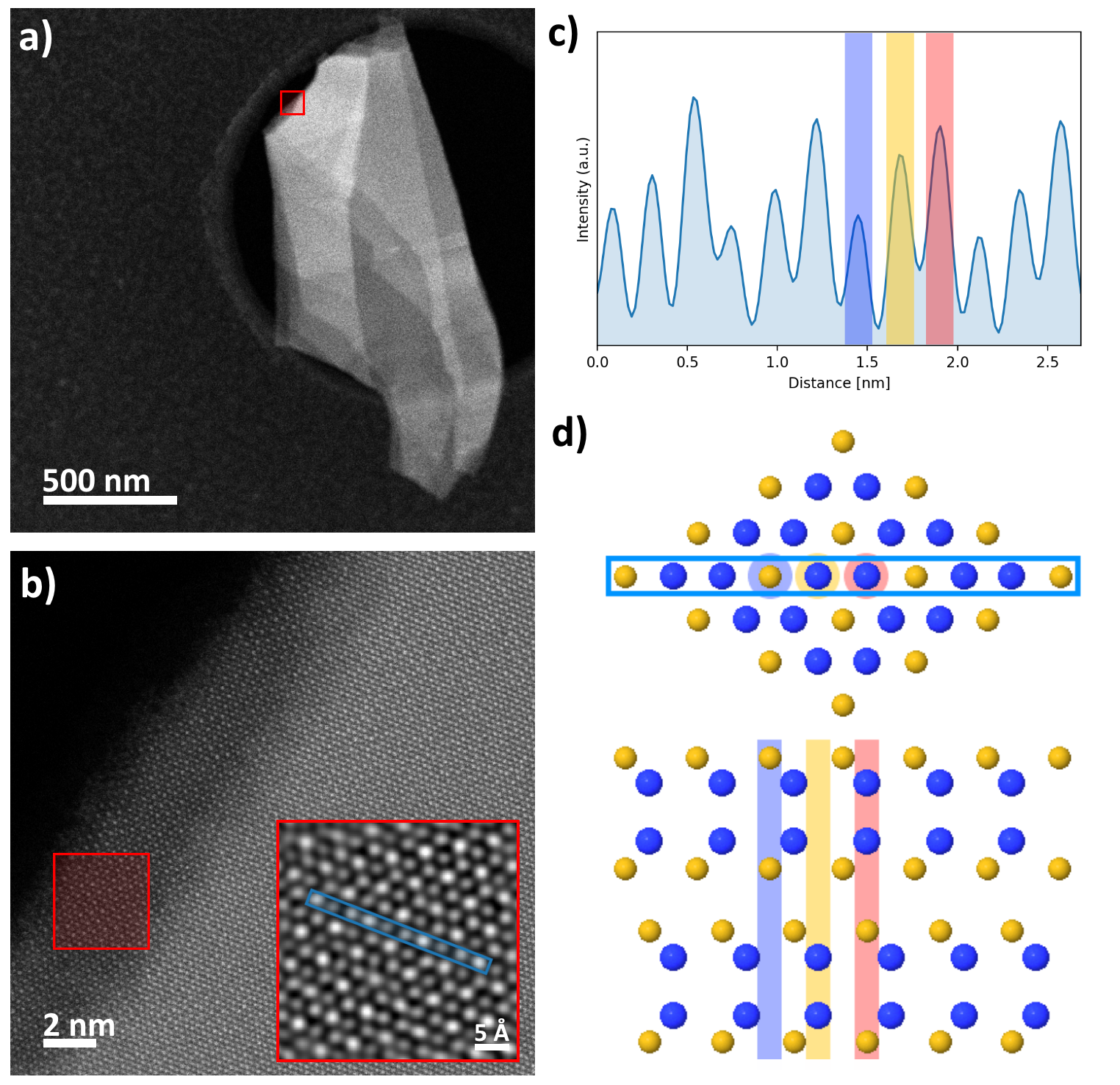}
  \vspace{0.2cm}
  \caption{\small (a) Low-magnification HAADF-STEM image of another $n$-doped InSe flake
    standing on top of a hole in the TEM grid.
    (b) High-resolution HAADF-STEM image acquired in the red square in (a),
    where the inset highlights the atomic crystalline structure.
    (c) HAADF intensity line profile taken along the blue
    rectangle in (b), where the three colors
correspond to the three-fold periodicity observed in the line profile.
    (d) Atomic model of $\varepsilon$-phase InSe, with the same
    choice of colors labeling the atomic layers as in (c), displaying
    top-view (upper) and cross-view (lower panel).
  }
\label{fig:FigSI-4}
\end{centering}
\end{figure}

In order to identify the crystalline phase of the specimen under consideration,
Fig.~\ref{fig:FigSI-4}(a)
displays a low-magnification HAADF-STEM image of a different InSe flake.
This flake has been
obtained from the same bulk material as that of Fig.~\ref{fig:FigSI-3}(a)
and hence shares its crystalline structure.
Notice how this  InSe flake is standing on top of a hole of the TEM grid.
Fig.~\ref{fig:FigSI-4}(b) then shows
a high-resolution HAADF-STEM image corresponding to the red square in (a),
and whose inset highlights the atomic arrangement in the region indicated with a red square.
Fig.~\ref{fig:FigSI-4}(c) provides the HAADF
intensity line profile taken along the blue
rectangle in (b), where the three colors
correspond to the three-fold periodicity observed in the line profile.
HAADF-STEM images are approximately proportional to $Z^{1.7}$, with $Z$ being the atomic number.
          By comparing with the expectations based on possible atomic models, these images
          provide useful information to identify the underlying crystalline sequence.
The line profile of Fig.~\ref{fig:FigSI-4}(c) is consistent with the
atomic model of $\varepsilon$-phase InSe shown in 
Fig.~\ref{fig:FigSI-4}(d), both for top-view and for cross-view,
and which uses the same
choice of colors as in Fig.~\ref{fig:FigSI-4}(c).

\begin{figure}[t]
\begin{centering}
  \includegraphics[width=0.80\linewidth]{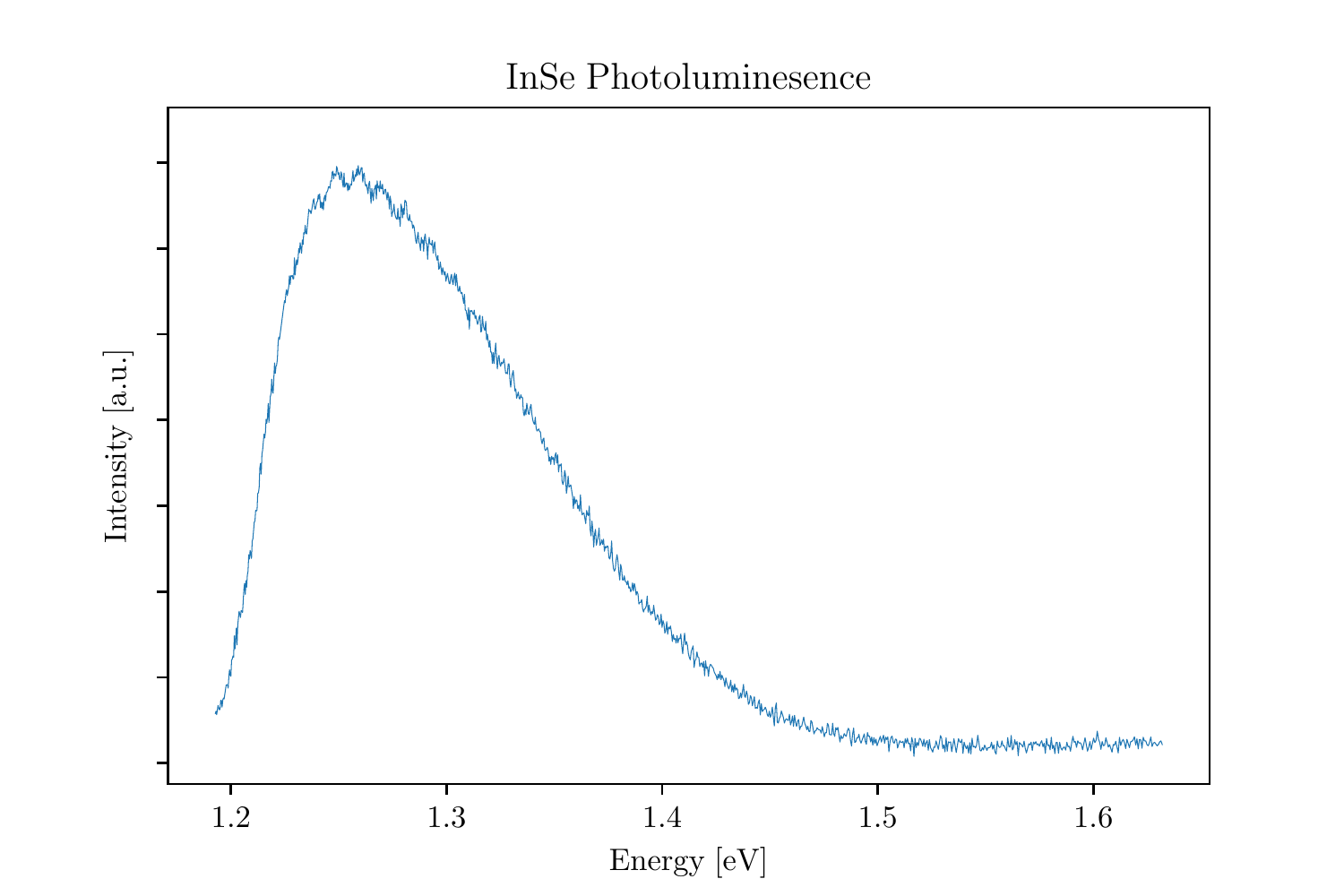}
  \vspace{0.2cm}
  \caption{\small Photoluminescence spectrum acquired
    in the $n$-doped InSe specimen.
    A well-defined peak, is observed
    indicating that this material is characterised
    by a direct bandgap with energy value $E_{\rm bg}\approx 1.26$ eV.
  }
\label{fig:Photo_luminesence}
\end{centering}
\end{figure}

The structural analysis presented in Fig.~\ref{fig:FigSI-4} indicates
that the $n$-doped InSe specimens considered in this work exhibit
a  crystalline structure characterised by a pure $\varepsilon$-phase.
In order to further elucidate the type of the bandgap exhibited by
this material, photoluminescence (PL) measurements
are carried out.
The results, displayed in Fig.~\ref{fig:Photo_luminesence},
exhibit a well-defined peak located around 1.26 eV.
Hence, we conclude that this material is characterised
by a direct bandgap with energy value $E_{\rm bg}\approx 1.26$ eV,
consistent with the findings of~\cite{henck2019evidence}.
Note that PL measurements are characterised
by a limited spatial resolution
as compared to the STEM-EELS results, and therefore this bandgap
value corresponds to an average across the specimen.
Hence, PL results are not sensitive to spatially-resolved
features in the bandgap map such as those reported
in Fig.~3(b) of the main manuscript.

\clearpage

\section{Bandgap analysis of 2H/3R WS$_2$ nanoflowers}
\label{sec:validation}

Here we apply our new approach to the bandgap analysis of same WS$_2$ specimen
considered in the original study~\cite{WS2_nanoflowers,Roest:2020kqy}.
This specimen consisted on a horizontally-standing WS$_2$ flake 
belonging to flower-like nanostructures characterised by a mixed 2H/3R polytypism.
While~\cite{Roest:2020kqy} restricted its bandgap analysis to a small subset
of individual EELS spectra, here we extend it to the whole specimen
and as a byproduct also we provide the local thickness map.
The goal is to demonstrate how our updated analysis is consistent
with the results presented in~\cite{Roest:2020kqy}.
The corresponding results for the dielectric function are presented
in App.~\ref{app:dielectric_ws2}.

\begin{figure}[ht]
\begin{centering}
  \includegraphics[width=0.99\linewidth]{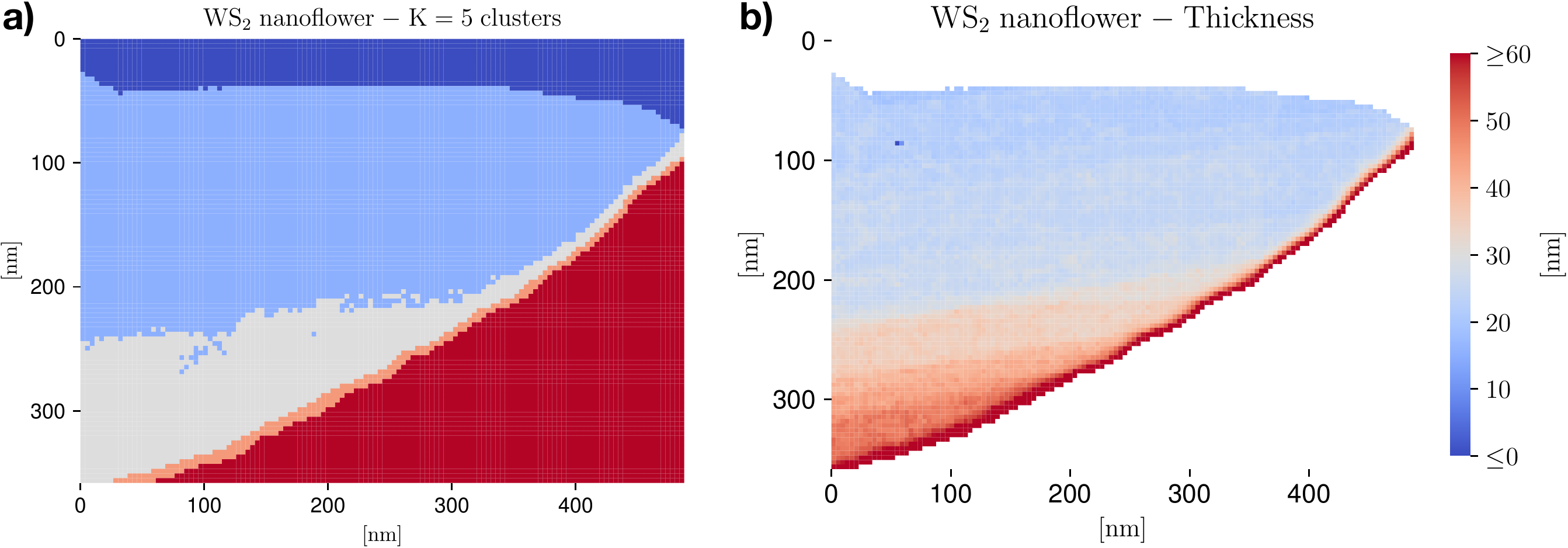}
  \vspace{0.2cm}
  \caption{Thickness analysis obtained from
    the EELS-SI presented in Fig.~5.1(a) of~\cite{Roest:2020kqy} as input.
    The specimen under consideration is  a horizontally-standing flake belonging to a
    2H/3R polytypic WS$_2$ flower-like nanostructure.
     (a) The outcome of the $K$-means clustering procedure for $K=5$.
    (b) The resulting thickness map. Only pixels associated to the WS$_2$ specimen are displayed.
  }
\label{fig:ws2_thickness}
\end{centering}
\end{figure}

Fig.~\ref{fig:ws2_thickness} displays the outcome of the thickness map
determination obtained using
the EELS-SI from Fig.~5.1(a) of~\cite{Roest:2020kqy} as input.
Fig.~\ref{fig:ws2_thickness}(a)
shows the results  of the $K$-means clustering procedure for $K=5$.
The choice of  $K=5$ clusters
is found to be a good compromise between minimizing the variance within each cluster
while ensuring a sufficiently large number of members,
as required by the applicability the Monte Carlo replica method.
The WS$_2$ specimen itself turns out to be classified into
three thickness clusters, surrounded by vacuum (dark blue) at the top
and by the SiN substrate (red) in the bottom-right region of the map.
Then Fig.~\ref{fig:ws2_thickness}(b) displays
the corresponding thickness map evaluated by means of
Eq.~(\ref{eq:thickness_calculation}).
Note that the image has been masked
by retaining only the pixels associated to the WS$_2$ specimen, to ease  visualization.
The qualitative agreement with the outcome of the $K$-means
clustering confirms
the reliability of the total integrated 
intensity $N_{\rm tot}$ as a suitable proxy for the local specimen thickness
when modelling the ZLP parametrisation.

\begin{figure}[ht]
\begin{centering}
  \includegraphics[width=0.99\linewidth]{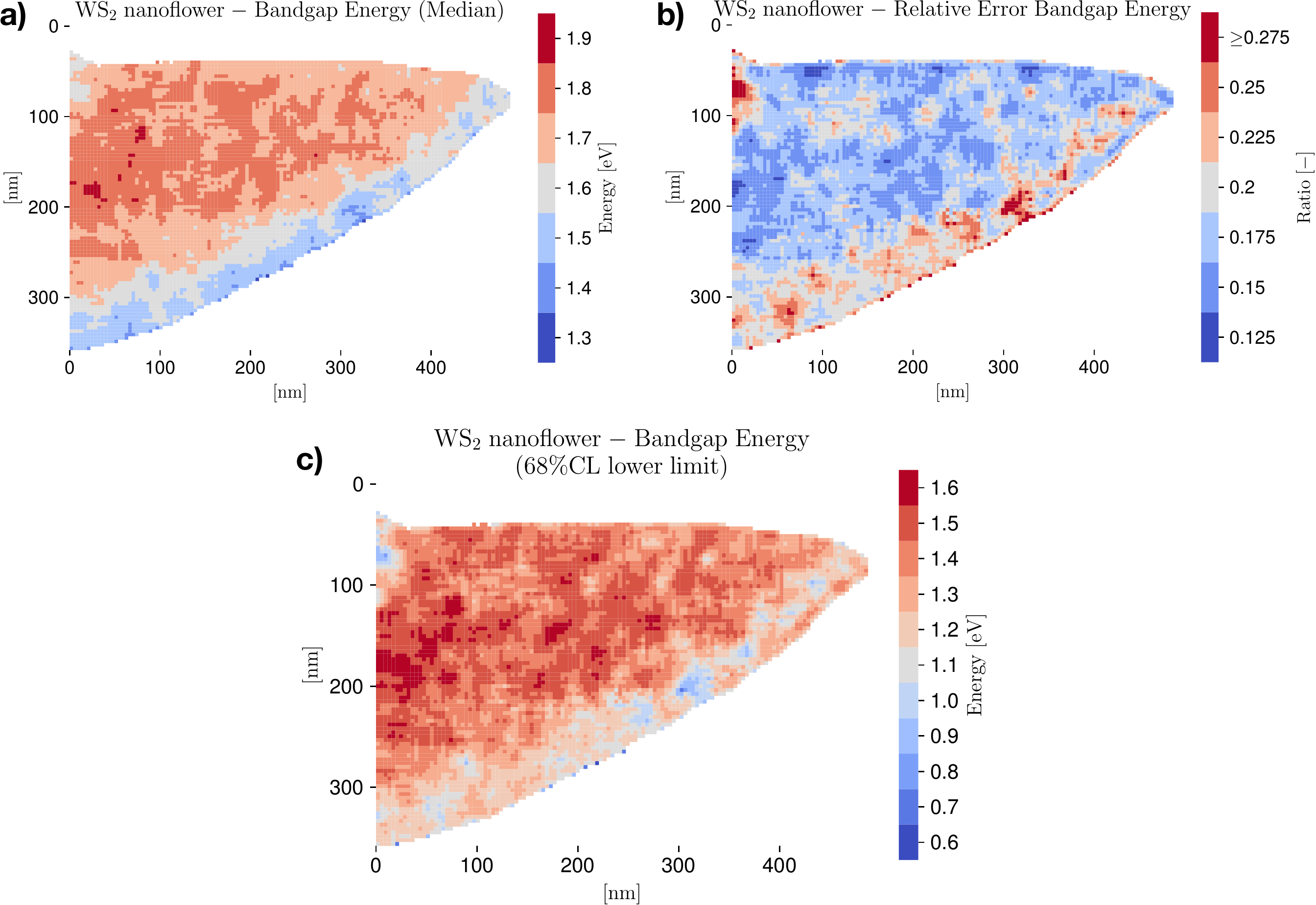}
  \vspace{0.2cm}
  \caption{Bandgap analysis of the same EELS-SI used to evaluate Fig.~\ref{fig:ws2_thickness}.
    (a) Spatially-resolved map displaying the median value of the  bandgap energy $E_{\rm bg}$
    evaluated across the specimen.
    (b) Same as (a) for the corresponding relative uncertainty $\delta E_{\rm bg}$, evaluated as
    half of the 68\% CL
    interval from the Monte Carlo replica sample.
    (c) Same  as (a) indicating the lower limit of the  68\% CL interval for $E_{\rm bg}$.
    In the three panels, the maps have been filtered such that only
    those pixels corresponding to the WS$_2$ specimen are retained.
  }
\label{fig:ws2_bandgap}
\end{centering}
\end{figure}

Fig.~\ref{fig:ws2_bandgap}
then presents the corresponding bandgap analysis obtained
from the same EELS-SI used to evaluate Fig.~\ref{fig:ws2_thickness},
where again the maps have been filtered such that only
    those pixels corresponding to the WS$_2$ specimen are retained.
First of all, Fig.~\ref{fig:ws2_bandgap}(a)
displays the spatially-resolved map displaying the median value of the  bandgap energy $E_{\rm bg}$
evaluated across the WS$_2$ specimen,  where
the spatial resolution achieved is around 10 nm.
These bandgap energies have been obtained from the procedure described in App.~\ref{sec:bandgap_analysis},
specifically  by fitting Eq.~(\ref{eq:I1}) to
the onset of the inelastic  spectra.
A fixed value of the exponent $b=1.5$, corresponding
to the indirect bandgap reported for this material, is used to stabilize the model fit.
Then Fig.~\ref{fig:ws2_bandgap}(b) shows the associated relative uncertainty $\delta E_{\rm bg}$
on the extracted bandgap energy.
It is estimated as half the magnitude of the 68\% CL
interval (corresponding to one standard deviation for a Gaussian
distribution) from the Monte Carlo replica sample for each pixel of the SI.
One finds that the typical uncertainties $\delta E_{\rm bg}$ range between 15\% and 25\%.
Finally, Fig.~\ref{fig:ws2_bandgap}(c) indicates the lower limit of the  68\% CL interval for $E_{\rm bg}$.

Ref.~\cite{Roest:2020kqy} reported
a value of the bandgap of 2H/3R polytypic WS$_2$ of $E_{\rm bg}=\lp
1.6\pm 0.3\rp $ eV
with a exponent of $b=1.3^{+0.3}_{-0.7}$ extracted from single EELS spectrum.
From the spatially-resolved bandgap maps of Fig.~\ref{fig:ws2_bandgap},
one observes how our updated results are in agreement with those from the previous study
within uncertainties.
Furthermore, this spatially-resolved determination of $E_{\rm bg}$ is
in agreement within uncertainties with first-principles calculations
based on Density Functional Theory (DFT) of the band structure of 2H/3R polytypic
WS$_2$~\cite{doi:10.1021/acsphyschemau.1c00038}.
These DFT calculations, which also account for spin-orbit coupling effects,
find values of $E_{\rm bg}$ in the range between 1.40 eV
and 1.48 eV depending on the settings of the calculation.
The DFT predictions are hence consistent with the 68\% CL interval for the  $E_{\rm bg}$ for a wide
region of the specimen, as indicated by  Fig.~\ref{fig:ws2_bandgap}(c).

Furthermore, inspection of the thickness and bandgap maps, Figs.~\ref{fig:ws2_thickness}(b)
and ~\ref{fig:ws2_bandgap}(a) respectively, reveals an apparent 
dependence of the value of $E_{\rm bg}$ on the local specimen thickness.
Specifically, the bandgap energies tend to increase in the thinner region
of the specimen, with $t\approx 25$ nm, and then to decrease as one moves towards
the thicker regions with $t\approx 50$ nm.
While this dependence with the thickness is suggestive of the known property of
WS$_2$ that $E_{\rm bg}$ increases when going from bulk to monolayer form, uncertainties
remain too large to be able to assign significance to this effect.

\clearpage

\section{Dielectric function in 2H/3R WS$_2$ nanoflowers}
\label{app:dielectric_ws2}

\begin{figure}[h]
\begin{centering}
  \includegraphics[width=0.99\linewidth]{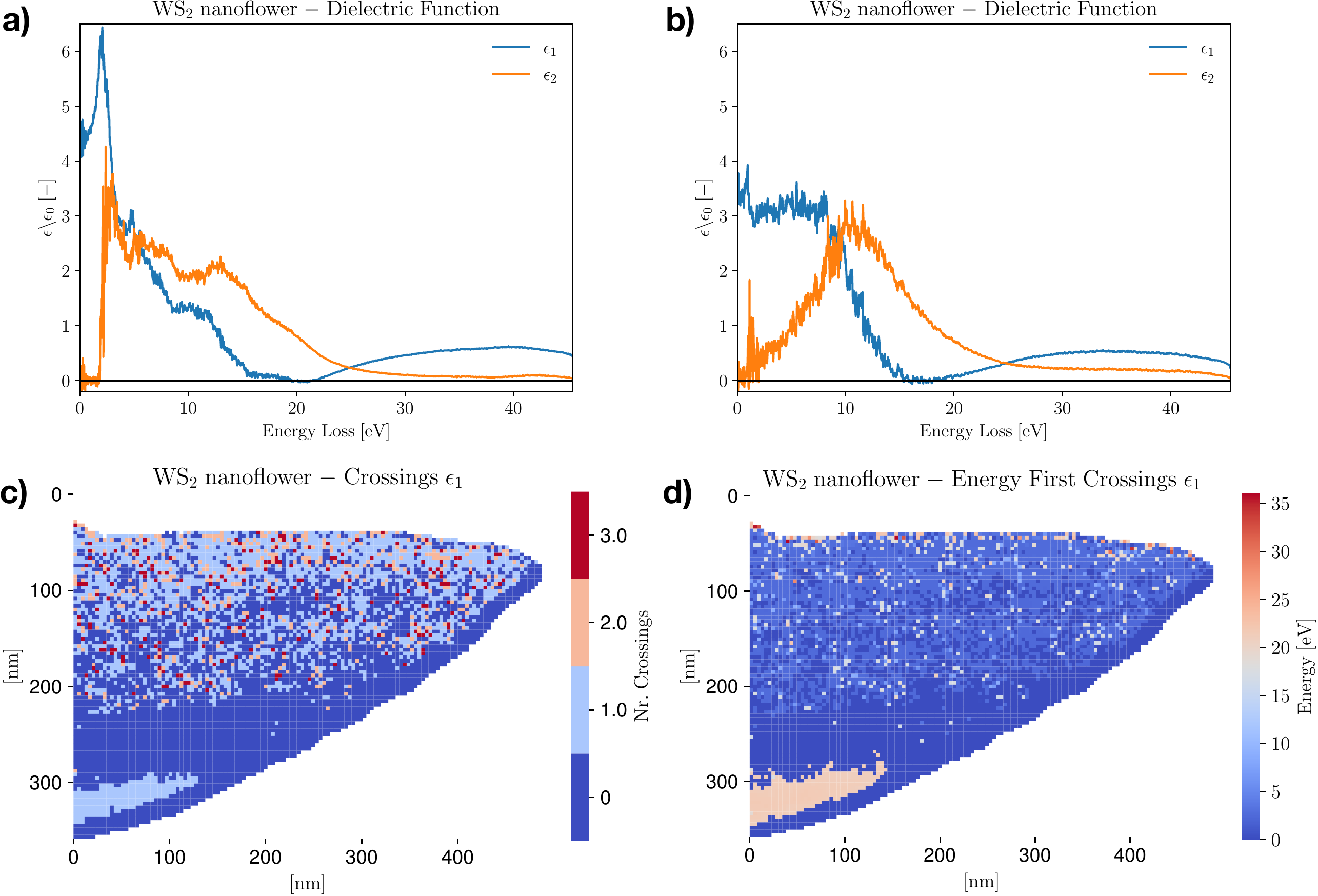}
  \vspace{0.2cm}
  \caption{(a,b) The real $\epsilon_1(E)$ and imaginary $\epsilon_2(E)$
    components of the complex dielectric
    function corresponding
    to two representative spectra of the WS$_2$ nanoflower specimen.
    The energy values for which the real component exhibits a crossing,
    $\epsilon_1(E)=0$ with positive slope, can be  attributed
    to collective electronic transitions.
    (c) The number of crossings
    exhibited by the real part
    of the dielectric function for the same specimen. 
    (d) The energy value of the left-most crossing
    in (c) for the pixels with $\ge 1$ crossings.
    Both in (c) and (d) only the  pixels corresponding to the specimen
    are retained.
    \label{fig:WS2_dielectric}
  }
\end{centering}
\end{figure}

App.~\ref{sec:validation}  characterizes the local thickness and the bandgap
energy of the 2H/3R WS$_2$ nanoflower specimen
from~\cite{WS2_nanoflowers,Roest:2020kqy} across the whole EELS-SI.
We now present the corresponding results for the spatially-resolved
determination of the real, $\epsilon_1(E)$, and imaginary, $\epsilon_2(E)$, parts
of its complex dielectric function.
Fig.~\ref{fig:WS2_dielectric}
displays $\epsilon_1(E)$
and $\epsilon_2(E)$  corresponding
to two representative spectra of this WS$_2$ nanoflower specimen.
In this analysis we account for the effects of the surface contributions and
the error bands quantify the uncertainties associated to the ZLP subtraction procedure.

Of particular interest  are the values of the energy loss 
for which the real component of the dielectric
function exhibits a crossing, $\epsilon_1(E)=0$ with a positive slope.
These crossings
can be interpreted as indicating a phase transition
involving a collective electronic excitation, such as a plasmonic resonance.
Here we define that a crossing takes place wherever
$\epsilon_1(E)=0$ at the 90\% CL as estimated
from the Monte Carlo representation.
For the selected spectra displayed in Fig.~\ref{fig:WS2_dielectric}(a,b),
this condition is satisfied for $E\simeq 22$ eV and $E\simeq 18$ eV
respectively.
These values are consistent with the bulk and surface  plasmonic
resonances in 2H/3R polytypic WS$_2$ identified in~\cite{WS2_nanoflowers}.

Fig.~\ref{fig:WS2_dielectric}(c) displays
the  number of such crossings exhibited by the real part
of the dielectric function across the whole WS$_2$ nanoflower specimen.
The vacuum and substrate regions have been masked such that only
the spectra corresponding to the specimen are retained.
One finds that the majority of the spectra are characterised by either one or zero
crossings, while a minority showing two or even three
crossings.
By comparing with Fig.~\ref{fig:ws2_thickness}(b), one observes how it is
in the thicker
region of the specimen for which the condition $\epsilon_1(E)=0$ is typically
not satisfied, with the exception of the very bottom region which consistently displays
one crossing. 
    
Finally,  Fig.~\ref{fig:WS2_dielectric}(d) displays the energy $E$
associated to the left-most crossing
in Fig.~\ref{fig:WS2_dielectric}(c) for those pixels with $\ge 1$ crossings.
For the upper region of the specimen (characterised by smaller thicknesses), the
first crossing is found to be in the low-loss region, while
in the bottom (thicker) region, one has a first crossing at $E_c\simeq 21$ eV
consistent with the WS$_2$ bulk plasmon peak.
We note that one could also show the values of $E_c$ for the subsequent crossings,
for those pixels exhibiting more than one crossing.

\clearpage

\section{Simultaneous determination of $(E_{\rm bg},b)$ from EELS-SI}
\label{sec:bandgap_analysis_exponent}

As discussed in the SI Sect.~\ref{sec:bandgap_analysis},
the ZLP-subtracted EELS spectra can be used to
extract the value of the bandgap energy $E_{\rm bg}$ as well as its type (direct or indirect)
in semiconductor  material by fitting a functional form
\begin{equation}
  \label{eq:I1_rep}
    I_{\rm inel}(E) \simeq  A \lp E-E_{\rm bg} \rp^{b} \, , \quad E \gsim E_{\rm bg} \, ,
\end{equation}
to the onset region of the subtracted inelastic spectra, where $b=0.5~(1.5)$
corresponds to a direct~(indirect) semiconductor.
In the results presented in the main manuscript, in particular in Fig.~3, fits based
on Eq.~(\ref{eq:I1_rep}) were carried out by fixing the value of the exponent $b$
to $b=0.5$ for InSe specimen, motivated by the PL analysis, and $b=1.5$ for the WS$_2$ nanoflower,
justified by previous work~\cite{WS2_nanoflowers,Roest:2020kqy} as well as by independent first-principles DFT calculations~\cite{doi:10.1021/acsphyschemau.1c00038}.
Here we present results of fits were both the bandgap energy $E_{\rm bg}$ and type $b$
are simultaneously determined from the ZLP-subtracted EELS spectra in the case of the InSe
specimen.

Already in our original study~\cite{Roest:2020kqy}, it was demonstrated
that our method is suitable for the joint  extraction of both
the bandgap energy $E_{\rm bg}$ and the exponent $b$
simultaneously, but that in general the latter is affected by sizable uncertainties.
For instance, for two of the EEL spectra considered there from the WS$_2$ specimen, we
found values $b=1.3^{+0.3}_{-0.7}$ and $b=1.3^{+0.3}_{-0.4}$,
consistent with the theoretical expectations for an indirect semiconductor.
One option to reduce the model uncertainties on the exponent parameter $b$ would be to increase the pooling degree of the pixels in the SI, which would reduce the fluctuations
  in the low-loss region at the price of a degradation of the achieved spatial resolution.
  Here instead we show how we can obtain important information about the values of the exponent $b$
  in a fully data-driven manner without compromising the spatial resolution.
  The strategy is to mask away the pixels where the joint determination of $(E_{\rm bg},b)$ is too noisy,
  and hence retain only those pixels where the relative uncertainties on
  the two parameters is below a precision threshold.
  
\begin{figure}[t]
\begin{centering}
  \includegraphics[width=0.49\linewidth]{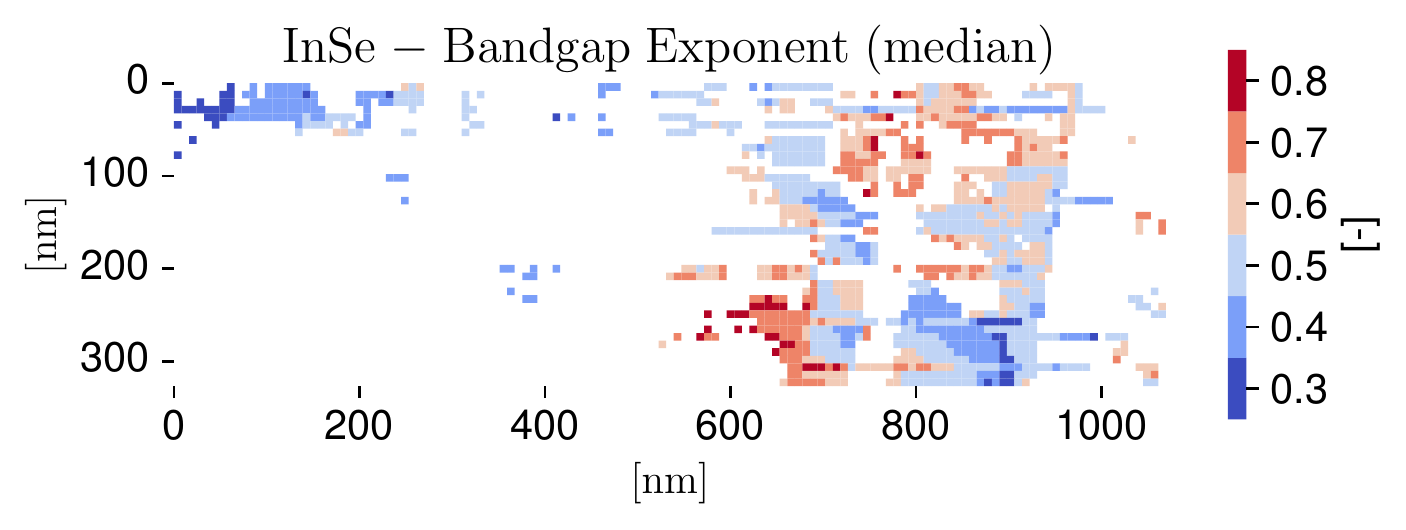}
  \includegraphics[width=0.49\linewidth]{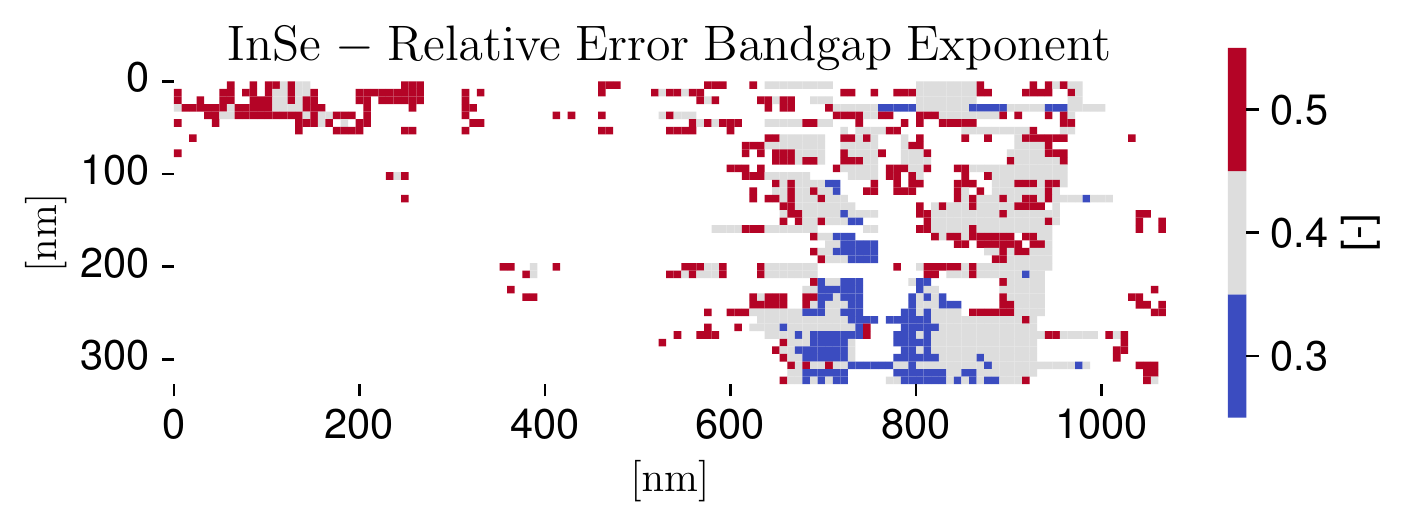}
  \includegraphics[width=0.49\linewidth]{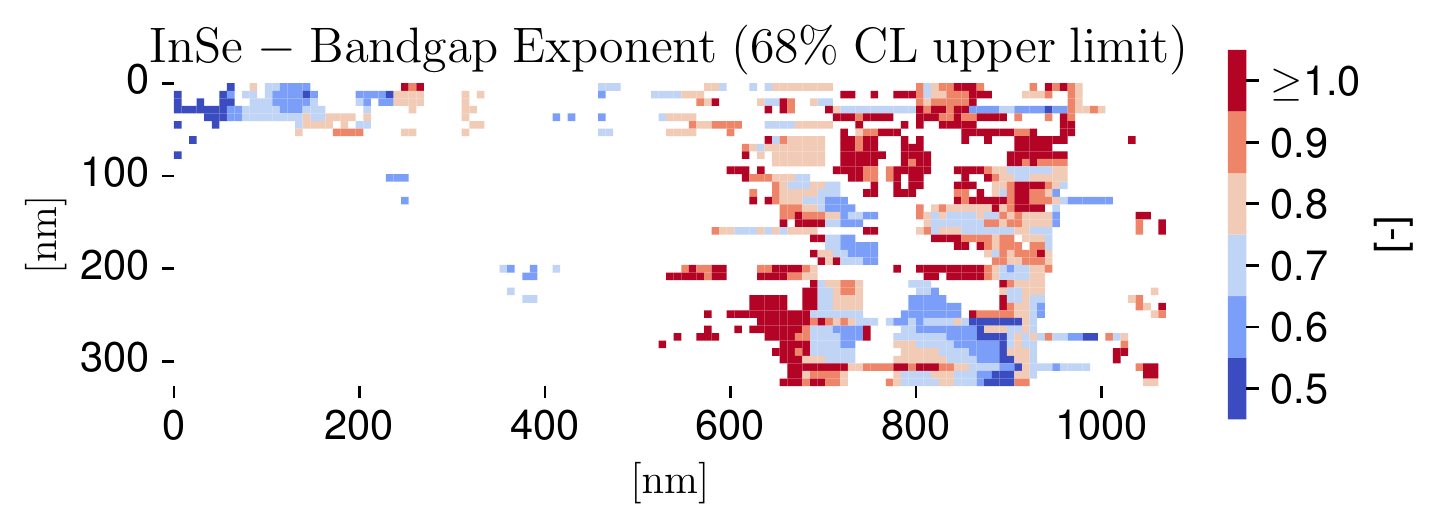}
  \includegraphics[width=0.49\linewidth]{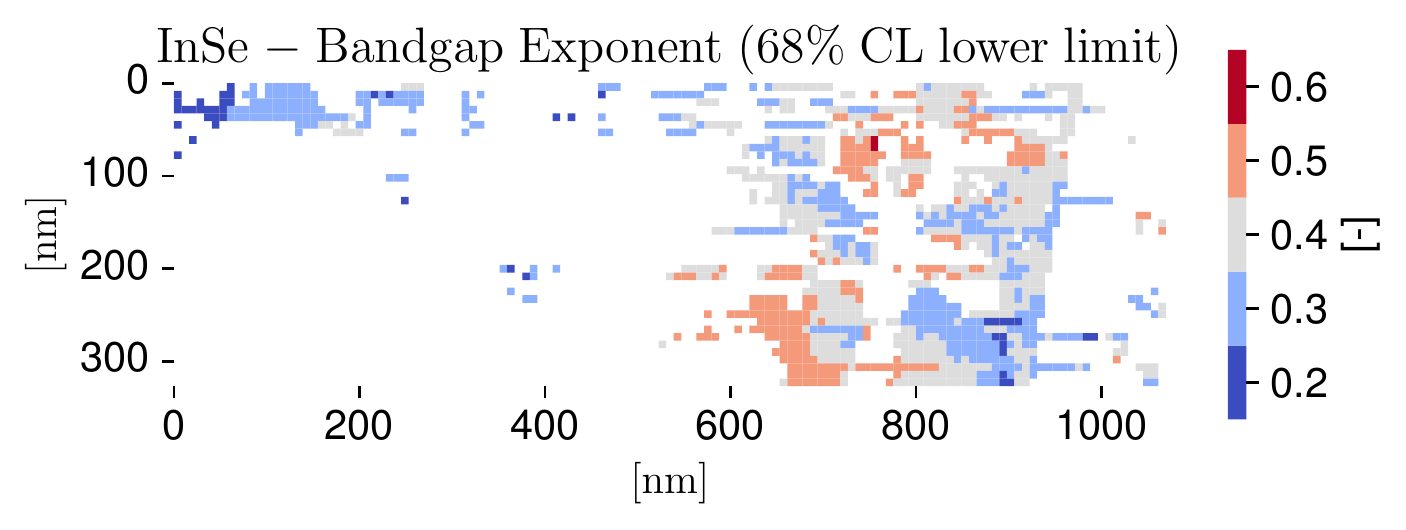}
  \vspace{0.2cm}
  \caption{Upper panels:  the median value (left) and 68\% CL relative uncertainty (right panel) for the exponent $b$
    in joint fits where the values of $(E_{\rm bg},b)$ are extracted simultaneously from the 
    low-loss region of the subtracted EELS spectra corresponding to the InSe specimen.
    Only those pixels where the relative uncertainty in the determination of both $b$ and $E_{\rm bg}$ is below
    the 50\% level are retained.
    Bottom panels: same, now corresponding to the upper (left) and lower (right panel) ranges of the 68\% CL
    interval for $b$.
  }
\label{fig:Ref1}
\end{centering}
\end{figure}
  
\begin{figure}[t]
\begin{centering}
  \includegraphics[width=0.49\linewidth]{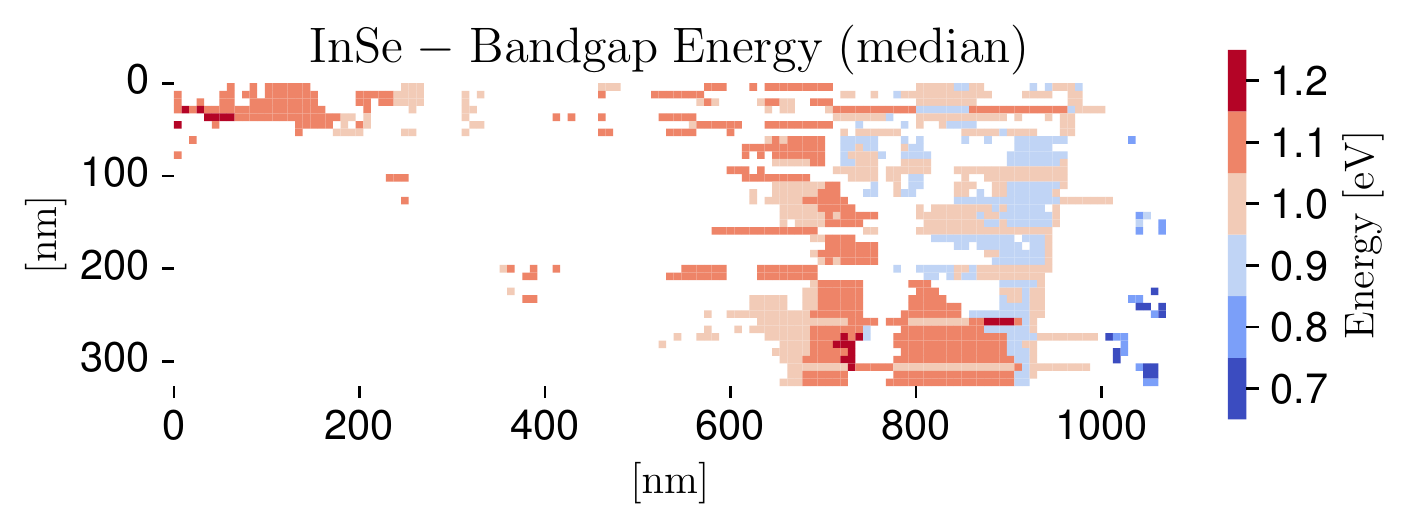}
  \includegraphics[width=0.49\linewidth]{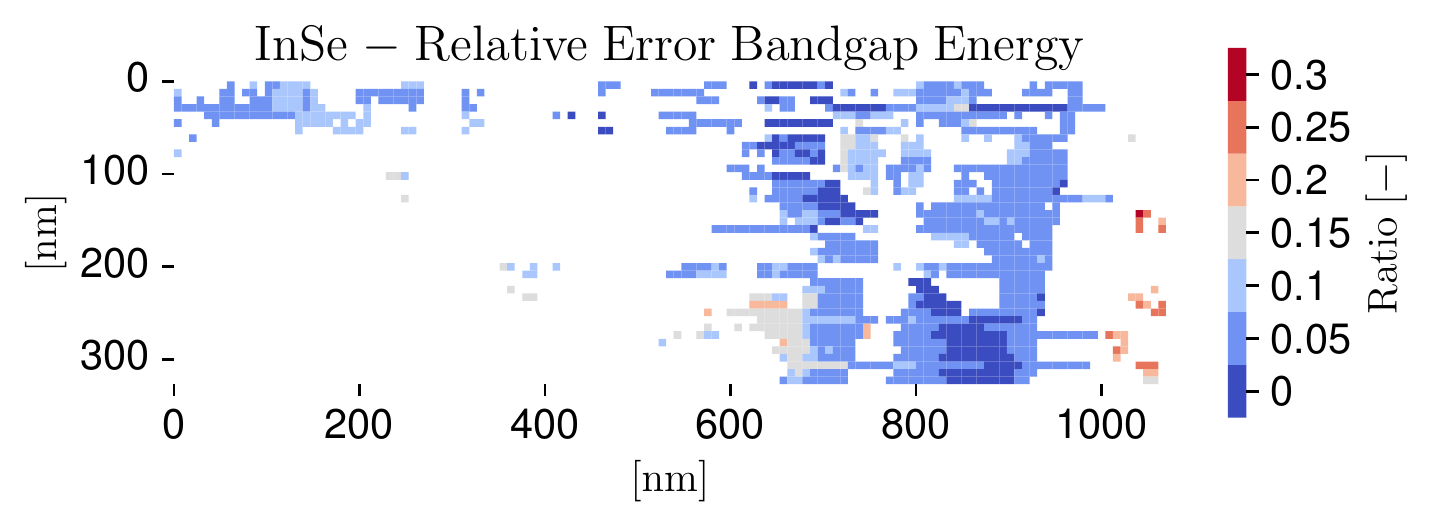}
  \includegraphics[width=0.49\linewidth]{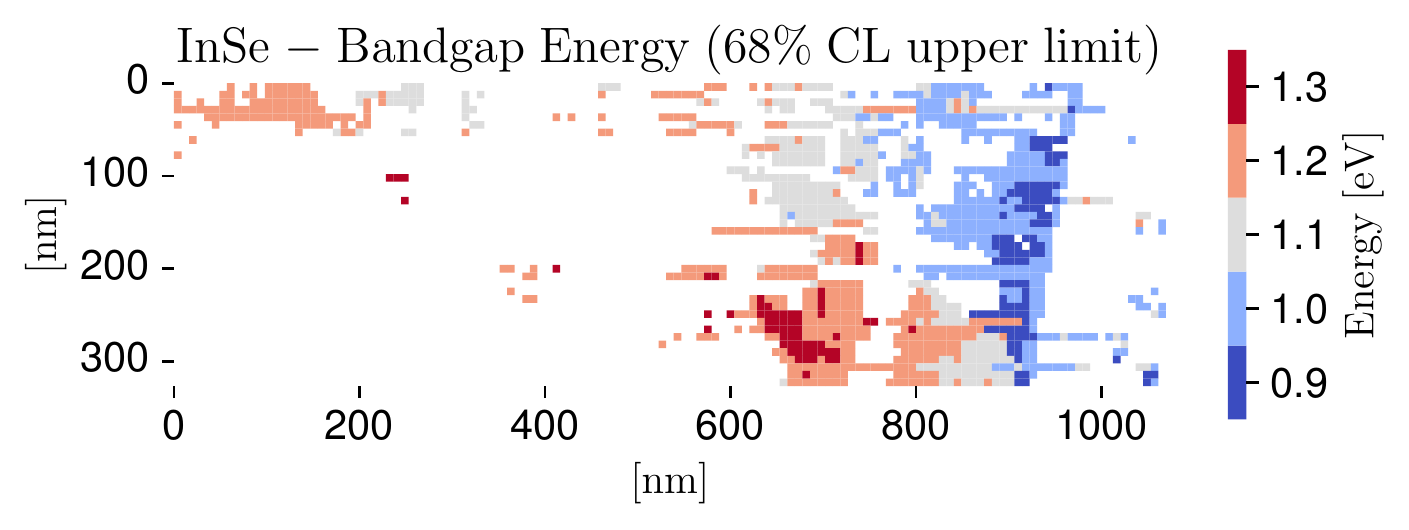}
  \includegraphics[width=0.49\linewidth]{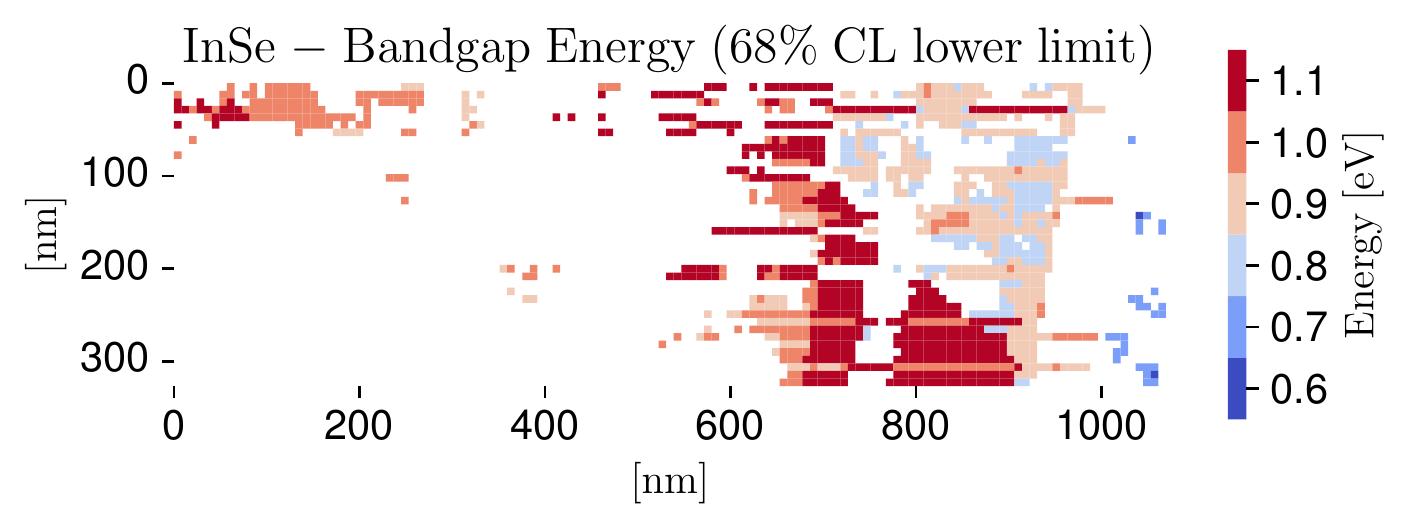}
  \vspace{0.2cm}
  \caption{Same as Fig.~\ref{fig:Ref1} now displaying the corresponding results
    for the bandgap energy $E_{\rm bg}$.
  }
\label{fig:Ref2}
\end{centering}
\end{figure}

  Figs.~\ref{fig:Ref1} and~\ref{fig:Ref2}
  display the outcome of joint fits of $(E_{\rm bg},b)$ where these two
  parameters are extracted simultaneously from the 
  low-loss region of the subtracted EELS spectra for the InSe specimen.
  The upper panels display the median value
  and 68\% CL relative uncertainty while the
  bottom panels show the values of the upper and lower ranges of the 68\% CL
  interval, in the case of the exponent $b$ and of the bandgap energy $E_{\rm bg}$ respectively.
  Only those pixels where the relative uncertainty in the determination of both $b$ and $E_{\rm bg}$ is below
  the 50\% level are retained, and the rest are  masked away.
  One finds that for around $25\%$ of the pixels it is possible to achieve
a reasonable precision in these joint fits of $(E_{\rm bg},b)$.

Inspection of Figs.~\ref{fig:Ref1} and~\ref{fig:Ref2} reveals that
the EELS data strongly prefers a value of the exponent around $b\simeq 0.5$,
consistent with both
the PL measurements and with the expectations of this material being a direct semiconductor,
while an alternative scenario with $b=1.5$ is strongly disfavored.
  Furthermore, one can verify that for all other pixels in the SI, including the ones masked away,
  $b=0.5$ is contained within the 68\% CL uncertainty range.
If we average over all the SI pixels displayed in Fig.~\ref{fig:Ref1} we find
$b = 0.50 \pm 0.26$, confirming that indeed the specimen is a direct semiconductor.
In addition, by comparing with Fig.~3~(b,c) of the main manuscript,
one finds that the extracted values of $E_{\rm bg}$ are stable
irrespectively of whether the exponent $b$ is kept fixed or instead  is fitted.

The analysis presented here demonstrates that our method is also suitable
to provide spatially-resolved information about the value of the bandgap
type, confirming the value of $b$ obtained from the  spatially-averaged PL data.
Further work will investigate how to improve the trade-off between
spatial resolution and precision in the joint fits of
$(E_{\rm bg},b)$ from the subtracted EELS spectra.


\providecommand{\latin}[1]{#1}
\makeatletter
\providecommand{\doi}
  {\begingroup\let\do\@makeother\dospecials
  \catcode`\{=1 \catcode`\}=2 \doi@aux}
\providecommand{\doi@aux}[1]{\endgroup\texttt{#1}}
\makeatother
\providecommand*\mcitethebibliography{\thebibliography}
\csname @ifundefined\endcsname{endmcitethebibliography}
  {\let\endmcitethebibliography\endthebibliography}{}

\end{document}